\documentclass[12pt]{article}
\usepackage[margin=1in]{geometry}
\usepackage{enumerate,enumitem}
\usepackage{listings}
\usepackage[usenames]{color}
\usepackage[table, svgnames, dvipsnames]{xcolor} % https://tex.stackexchange.com/q/373436
\usepackage{url,graphicx,tabularx,array}
\usepackage{adjustbox}
\usepackage{pdflscape} %\begin{landscape} \end{landscape}
\usepackage{multirow}
\usepackage{amsmath,amsthm,amsgen,amstext,amsbsy,amsopn,amsfonts,amssymb} 
\usepackage[hyperindex=false]{hyperref}
\usepackage[noabbrev, capitalise]{cleveref}
\AtBeginEnvironment{appendices}{
	\crefalias{section}{appendix}
	\crefalias{subsection}{appendix}
} % https://tex.stackexchange.com/a/121055
\usepackage{makeidx}
\usepackage[titletoc,title]{appendix}
\usepackage{chngcntr} % Table IA
\usepackage{titlesec}
\usepackage[utf8]{inputenc}
\usepackage{booktabs,dcolumn}
\usepackage{etoolbox} % https://tex.stackexchange.com/a/325698 % parskip; blockquote
\usepackage[nodisplayskipstretch]{setspace} % \doublespacing; comment by Ivanov: https://tex.stackexchange.com/a/224989
\usepackage{titling} % move title and authors up https://code.whatever.social/exchange/tex/questions/29593/shift-title-and-author-text-up
\usepackage[bottom]{footmisc} %footnotes at the bottom
\usepackage{caption, subcaption} % https://stackoverflow.com/a/51490189
\usepackage{makecell} % \makecell{a \\ b} https://tex.stackexchange.com/a/176780
\usepackage{float} % H for tables and figures staying here; but don't use [H] around landscape: https://tex.stackexchange.com/a/383889
\usepackage{authblk} % authors and affiliations: https://tex.stackexchange.com/a/214408
\usepackage{doc} % https://tex.stackexchange.com/questions/228807/thanks-footnote-indented-when-using-amsart-and-doc

\makeatletter
\def\@setthanks{\vspace{-\baselineskip}\def\thanks##1{\@par##1\@addpunct.}\thankses}
\makeatother

\makeatletter % https://tex.stackexchange.com/a/2652
\g@addto@macro\@floatboxreset\centering
\makeatother

\setkeys{Gin}{width=\textwidth, height=\textheight, keepaspectratio} % https://tinyurl.com/pandoc-tex-graph
\providecommand{\tightlist}{%
  \setlength{\itemsep}{0pt}\setlength{\parskip}{0pt}}
\makeatletter
\newsavebox\pandoc@box
\newcommand*\pandocbounded[1]{% scales image to fit in text height/width
  \sbox\pandoc@box{#1}%
  \Gscale@div\@tempa{\textheight}{\dimexpr\ht\pandoc@box+\dp\pandoc@box\relax}%
  \Gscale@div\@tempb{\linewidth}{\wd\pandoc@box}%
  \ifdim\@tempb\p@<\@tempa\p@\let\@tempa\@tempb\fi% select the smaller of both
  \ifdim\@tempa\p@<\p@\scalebox{\@tempa}{\usebox\pandoc@box}%
  \else\usebox{\pandoc@box}%
  \fi%
}

\hypersetup{
	colorlinks=true,
	linkcolor=blue,
	filecolor=magenta,
	urlcolor=blue,
	citecolor = blue
}

\usepackage[authordate, maxcitenames=2, backend=bibtex]{biblatex-chicago}
\title{ % \singlespacing
	Do Customer Disclosures Affect Suppliers' Internal Capital Allocation Decisions? % Evidence from SFAS 131
}

\ifdefined\blindsubmission
\author{}
\hypersetup{
	pdftitle={Do Customer Disclosures Affect Suppliers' Internal Capital Allocation Decisions?},
	pdfauthor={}
}
\else
\author{ % \singlespacing
Sangwook Nam\thanks{
	E-mail: \href{mailto:}{s.nam@binghamton.edu}\\
	I am grateful to my dissertation committee members: Sugata Roychowdhury (chair), Ron Dye, Jung Min Kim, and Mihir Mehta for their guidance and feedback. I also thank Regina Wittenberg-Moerman, Andy Leone, Beverly Walther, Aaron Yoon, Georg Rickmann, Tom Hagenberg, Jen Choi, Furkan Cetin, Yongseok Kim, Danling Song (discussant), Ying Zhou, Chongho Kim, Gary Chen, Hojun Seo, Jaewoo Kim (discussant), In Gyun Baek, Seung Hyeong Lee, Brianna de la Osa (discussant), Dirk Black, and workshop participants at the Kellogg School of Management, the 2024 Annual WashU Olin Accounting Research Conference (PhD poster session), the 2024 Boston Accounting Students Symposium, the 2024 KAAPA PhD Conference, the 2025 Hawaii Accounting Research Conference, the 2026 International Conference of the Journal of International Accounting Research (JIAR), CUHK--Shenzhen, and Binghamton University. I thank Young Jun Cho for sharing a list of firms that changed segment disclosures due to the adoption of SFAS 131. I thank Alex Sung (Samsung Electronics) and Steve Arntson (Merck) for their insights. I am grateful for the financial support from Kellogg School of Management. All errors are my own.}}
\affil{Binghamton University}
\fi

\date{August 26, 2026}

\begin{document}
\clearpage\maketitle
\thispagestyle{empty}

\begin{abstract} \singlespacing \noindent
	This study examines whether customer disclosures affect how supplier
firms allocate capital across business segments. Customer disclosures
can shape supplier investment decisions through two competing channels.
They can improve suppliers' information about downstream demand, helping
suppliers align capital with growth opportunities (``information
channel''), or erode incumbent suppliers' private information advantage,
inducing costly investments to defend customer relationships
(``competitive-threat channel''). I use the adoption of SFAS 131 as a
customer-level disclosure shock. Suppliers exposed to expanded customer
disclosures experience increased product-market competition and
reallocate capital toward segments with relatively weak
growth-opportunity signals. Suppliers that deviate from allocations
predicted by growth signals are more likely to preserve market share and
expand their customer base in subsequent years. Using a novel approach
to link supplier segments to customer segments, I show that this
reallocation is driven by investing in capacity in affected segments,
rather than by a correction of prior under-investment. Consistent with
the competitive-threat channel, the investment adjustment is stronger
for segments linked to larger customers and for segments operating in
more concentrated industries. Segments making these investments
subsequently experience lower ROA, consistent with suppliers accepting
lower profitability to defend customer relationships. Overall, the
findings show that disclosures can shape how economically linked firms
allocate capital internally.

\end{abstract}



\newpage
\doublespacing
\setcounter{page}{1}

\section{Introduction}\label{introduction}

A supplier firm does not make investment decisions in isolation. These
decisions depend on its growth opportunities and on information about
the prospects of its customers. For suppliers operating multiple
business segments, this information affects not only whether to invest,
but also where \emph{within the firm} to invest. Because capital is
scarce, allocating more resources to one business segment leaves fewer
resources available for others. Internal capital allocation is therefore
central to understanding how customer disclosures shape suppliers'
investment decisions.

A large literature studies how internal capital markets allocate
resources across segments. Existing studies explain internal capital
allocation through either frictions within the firm
\autocite{stein-1997-internal,scharfstein-stein-2000-dark} or external
shocks that change the firm's resource availability
\autocite{lamont-1997-cash}. However, less is known about whether
information disclosed by customers affects how suppliers allocate
internal capital across segments. To address this gap, this paper
examines whether customer disclosures induce suppliers to allocate
internal capital toward growth opportunities or instead reallocate
capital in response to competitive threats from potential suppliers.

Customer disclosures may reshape suppliers' internal capital allocation
decisions through two conceptually distinct channels. On the one hand,
customer disclosures may reveal downstream demand and growth
opportunities that the supplier could not otherwise observe (the
``information channel''). If uncertainty about customers constrains
relationship-specific investment, expanded disclosures should help the
supplier correct prior under-investment by directing capital toward the
supplier's highest-growth-opportunity segments.

On the other hand, customer disclosures may erode the incumbent
supplier's information advantage (the ``competitive-threat channel'').
Supplier-customer relationships often generate private knowledge about
customer needs and segment-level demand. Once disclosed publicly, that
information can help potential suppliers identify profitable investment
opportunities and target the incumbent supplier's customer base. The
incumbent supplier may respond by committing capacity to the affected
segments to protect customer relationships, even when standard growth
signals do not justify the additional investment
\autocite{spence-1977-entry,huisman-kort-2015-strategic,cookson-2018-anticipated}.
Distinguishing between these channels requires examining both where
suppliers allocate capital internally and whether those capital
allocations are consistent with improved growth opportunities or with
strategic responses to competitive threats.

A key empirical challenge is to identify a plausibly exogenous change in
customer disclosures. This paper addresses this challenge by using the
adoption of SFAS 131 as a quasi-natural experiment. Effective for fiscal
years beginning on or after December 15, 1997, SFAS 131 replaced SFAS
14's industry-based segment reporting framework with the management
approach. Under SFAS 131, firms report business segments in a manner
consistent with management's internal decision-making structure. For
firms that changed their segment reporting (henceforth, \emph{``change
firms''}), the standard expanded the public availability of
segment-level information on sales, capital expenditures, assets, and
profits.

Because the expanded disclosures pertain to particular customer
segments, a second empirical challenge is to determine which supplier
segments are economically exposed to the newly disclosed information.
This study addresses this challenge by using Bureau of Economic Analysis
(BEA) Input-Output tables to link customer segments to supplier segments
based on industry-level production relationships. This mapping
translates customer-level disclosure changes into
supplier-segment-specific measures of information exposure. This mapping
therefore allows me to examine whether suppliers reallocate capital
toward the segments most exposed to the expanded customer
disclosures.\footnote{See \cref{sec:data_collection} for details of the
  BEA-based segment-mapping procedure.}

Prior research establishes that SFAS 131 changed both the
informativeness and the competitive implications of segment disclosures.
Studies find that the standard increased the transparency of change
firms by providing more disaggregated information and making the
performance of individual segments visible to outsiders
\autocite{berger-hann-2003-impact,berger-hann-2007-segment,cho-2015-segment}.
At the same time, the additional disaggregation may give potential
suppliers access to information that incumbent suppliers could
previously obtain through their customer relationships. Consistent with
this proprietary concern, managers had substantial discretion under SFAS
14 to aggregate activities into broadly defined industry segments
\autocites[e.g.,][]{harris-1998-association,botosan-stanford-2005-managers,zhou-2022-proprietary}.
Thus, SFAS 131 provides a setting in which expanded customer disclosures
may both improve incumbent suppliers' information about segment-level
investment opportunities and reduce their information advantage over
potential suppliers. I distinguish between these two channels by
examining how capital allocations vary across supplier segments with
different exposure to the expanded disclosures and whether those
allocations align with underlying growth opportunities.

To implement this analysis, I construct a sample spanning 1996 to 2000,
centered on the adoption of SFAS 131 in 1998. For up to two pre-adoption
years, I use restated data from change firms' first post-adoption 10-Ks
to measure segment capital expenditures under consistent segment
definitions. The narrow window also limits mechanical changes in segment
composition arising from subsequent restructurings, divestitures, or
acquisitions.\footnote{Following \textcite{berger-hann-2003-impact} and
  \textcite{cho-2015-segment}, I exclude suppliers that experience
  restructuring, divestitures, or mergers and acquisitions to reduce the
  possibility that operational changes drive the investment results.}
Following adoption, 43\% of customer firms in the sample changed their
segment disclosures.\footnote{The remaining 57\% of customer firms
  either were single-segment firms or had voluntarily disclosed segments
  consistently with SFAS 131.} Of those change firms, 92\% reported more
segments under SFAS 131 than under SFAS 14.\footnote{The remaining 8\%
  of change firms either retained the same number of segments but
  reconfigured them or reported fewer segments. I exclude these firms
  from the set of customer-disclosure shocks.} I use these expanded
disclosures as the customer-level disclosure shock.

I first examine whether customer disclosures change the competitive
environments faced by affected suppliers. Using Text-based Network
Industry Classifications (TNIC) measures
\autocite{hoberg-phillips-2010-product,hoberg-phillips-2016-textbased},
I find that suppliers with greater exposure to expanded customer
disclosures experience both an increase in the number of rivals and
greater textual similarity to those rivals. These results are consistent
with potential suppliers moving closer to incumbent suppliers' product
space as customers expand their disclosures.

Next, I use a difference-in-differences (DiD) design to examine how
expanded customer disclosures affect suppliers' internal capital
allocation decisions. Treatment intensity, \(\% Treat\), is defined as
the share of a supplier's major customers that expand segment
disclosures following SFAS 131.\footnote{I measure the size of a
  supplier's major-customer base by the number of customers disclosed in
  the supplier's 10-K. The number of these customers forms the
  denominator of the treatment-intensity measure.} The dependent
variable is a firm-level measure of whether capital is allocated toward
segments with stronger or weaker benchmark growth opportunities
\autocite{cho-2015-segment}. Higher values indicate a relative shift
toward segments with stronger benchmark growth opportunities, whereas
lower values indicate a shift toward segments with weaker opportunities.

This paper finds that treated suppliers reallocate capital toward
segments that appear weak relative to their stronger sibling segments.
This result is less consistent with the information channel, which
predicts that expanded customer disclosures should lead to closer
alignment between capital allocations and benchmark growth
opportunities. If this reallocation reflects a strategic response to
competitive threats, it should help suppliers preserve or strengthen
their market position, consistent with the competitive-threat channel.

The subsequent firm-level outcomes are consistent with this
interpretation. I find that suppliers that deviate from the capital
allocation prescribed by growth signals are more likely to preserve
market share and expand their customer base in subsequent years. These
results can be understood as reflecting a trade-off between the cost of
expanding capacity and the benefit of retaining customers. To the extent
that the expected loss from customer attrition exceeds the cost of
capacity expansion, reallocating capital toward affected segments is a
rational response to increased competitive pressure. These findings
therefore suggest that the reallocation serves a strategic objective.
Although these findings clarify the strategic consequences of the
reallocation, the firm-level internal capital allocation measure does
not distinguish whether this relative shift is driven by increased
investment in weaker segments or reduced investment in stronger
segments.

To distinguish between the two underlying sources of the firm-level
result, I evaluate investment at the supplier-segment level by extending
the sales-growth model of investment \autocite{biddle-etal-2009-how}. I
define treated segments as supplier segments economically linked to
customers that expanded their segment disclosures.\footnote{The results
  remain similar when I alternatively define treated segments as
  supplier segments linked to customer segments newly disclosed
  following SFAS 131.} Using the BEA-based mapping described above, I
regress each supplier segment's lead investment on its own sales growth
and the sales growth of the linked customer segment. The residual from
the model captures investment that is high or low relative to these
observable growth signals.

The segment-level results indicate that the reallocation is driven by
increased investment in weaker segments. Treated segments have more
positive investment residuals, suggesting that they invest more than
their own and their linked customers' growth signals would predict. By
contrast, the incidence of negative residuals does not change. This
pattern is more consistent with the competitive-threat channel and less
consistent with the information channel, under which expanded customer
disclosures should help suppliers reduce negative investment residuals
where observable growth signals justify greater investment.

To further distinguish the competitive-threat channel from the
information channel, I perform two cross-sectional tests and one
consequence test. The cross-sectional tests examine two settings in
which supplier segments have stronger incentives to respond to increased
competitive pressure. First, the positive investment residuals increase
more for affected supplier segments linked to larger customers. This
result is less consistent with the information channel because
information frictions are likely to be greater for smaller customers. If
customer disclosures primarily reduce those frictions, the investment
response should be stronger for supplier segments linked to smaller
customers. Second, the positive investment residuals increase more for
affected supplier segments operating in more concentrated industries,
where incumbents have greater market power and thus greater rents at
risk from increased competition. These findings are consistent with the
competitive-threat channel, which predicts stronger investment responses
when customer relationships or incumbent rents are more valuable.

I next examine the subsequent profitability of the investments captured
by these positive investment residuals. If expanded customer disclosures
primarily alleviate uncertainty and information frictions, the resulting
investments should reflect a more efficient allocation of capital and
therefore should not be followed by lower operating performance. By
contrast, the competitive-threat channel predicts suppliers would
undertake costly strategic investments to protect customer relationships
or deter entry by potential suppliers, even at the expense of
profitability. Consistent with the competitive-threat channel, supplier
segments with positive investment residuals experience lower subsequent
ROA, suggesting that they accept lower profitability to defend important
customer relationships and incumbent market positions. Taken together,
the cross-sectional and consequence tests provide further support for
the competitive-threat channel over the information channel.

In addition to these segment-level investment results, suppliers with
greater exposure to expanded customer disclosures also increase
advertising and SG\&A spending at the firm level. These findings suggest
that suppliers' responses to expanded customer disclosures extend beyond
capital investment to other expenditures consistent with heightened
competitive threats \autocite{smiley-1988-empirical}.

A key identifying assumption of the DiD design is that suppliers with
different treatment exposure would have followed parallel investment
trends in the absence of the disclosure shock. I find neither
differential pre-trends in the firm-level or segment-level investment
measures nor differences in pre-mandate average investment between the
high- and low-\(\% Treat\) groups for the firm-level measure or between
treated and control supplier segments for the segment-level measure.
These tests mitigate the concern that the estimates capture a
pre-existing investment trend that becomes visible around the mandate.
More generally, for an omitted factor to explain the results, it would
need to be associated with suppliers' exposure to customers that
expanded their segment disclosures and generate differential changes in
supplier investments around the implementation of SFAS 131.

An alternative explanation is that affected suppliers reallocate capital
because customer disclosures heighten competition in their customers'
industries, rather than because those disclosures expose suppliers
themselves to greater competitive threats. Under this explanation,
greater competition among customers could prompt suppliers to reallocate
capital to support their customers' responses to the increased
competition. The evidence does not support this explanation. Customers
that expand their segment disclosures experience no statistically
significant increase in either the number of rivals or their textual
similarity to those rivals. Moreover, changes in customers' competitive
environments do not appear to alter how suppliers reallocate internal
capital in response to customer disclosures.

This paper contributes to several streams of the literature. First, it
extends research on the spillover effects of disclosure. Prior studies
show that disclosures influence firms' investment decisions, but the
evidence is largely at the firm level
\autocite{raman-shahrur-2008-relationshipspecific,badertscher-etal-2013-externalities,shroff-etal-2014-information,chiu-etal-2019-customers,durnev-mangen-2020-spillover,afrin-etal-2025-internalizing}.
I extend this literature by showing that customer disclosures affect not
only suppliers' aggregate investment decisions, but also their internal
capital allocation decisions across segments.

Second, this paper contributes to the literature on internal capital
allocation. Existing research emphasizes internal determinants of
allocation decisions, including winner-picking, financing constraints,
agency conflicts, and information frictions within the firm
\autocite{stein-1997-internal,billett-mauer-2003-crosssubsidies,duchin-sosyura-2013-divisional,chen-etal-2018-clarity}.
I identify an external determinant by showing that the competitive
information environment created by customer disclosures also shapes
headquarters' allocation decisions.

Third, this study extends research on segment reporting. Prior studies
of SFAS 131 show that the standard increased segment transparency and
improved monitoring of disclosing firms
\autocite{berger-hann-2003-impact,botosan-stanford-2005-managers,ettredge-etal-2005-impact,berger-hann-2007-segment,cho-2015-segment}.
I broaden this literature by documenting that segment disclosures also
affect the investment decisions of firms connected to the discloser
through supply-chain relationships.

The remainder of the paper develops the hypotheses in
\cref{sec:bgrd_hypo}, describes the data and sample-selection process in
\cref{sec:data_sample}, and presents the research design in
\cref{sec:research_design}. \cref{sec:emp_results} reports the main
results, and the paper concludes in \cref{sec:conclusion}.

\section{Background and Hypothesis Development}\label{sec:bgrd_hypo}

\subsection{SFAS 131 and Segment-Disclosure
Transparency}\label{sfas-131-and-segment-disclosure-transparency}

SFAS 131 changed how public firms report their business segments. Under
the previous standard, SFAS 14, firms reported segments using an
industry approach that gave managers considerable latitude to combine
operations with different economic characteristics
\autocite{herrmann-thomas-2000-analysis}. SFAS 131 replaced that
approach with the management approach, under which firms report segments
in a manner consistent with management's internal decision-making
structure \autocite{fasb-1997-statement}. By aligning segment reporting
with firms' internal organizational structures, the standard was
intended to improve the relevance of segment disclosures and enhance
external monitoring and transparency.

SFAS 131 produced an increase in public information about firms that
expanded their segment reporting upon adoption (hereafter, ``change
firms''). Comparisons of historical SFAS 14 disclosures with restated
SFAS 131 disclosures show that the new standard increased the number of
reported segments and revealed information about operations that had
previously been aggregated \autocite{berger-hann-2003-impact}. For
customers that expanded their segment disclosures, the mandate therefore
made segment-level performance and business composition more visible to
external stakeholders, e.g., suppliers and competitors.

Proprietary costs help explain why this information was not fully
disclosed before the mandate. When disclosure can reveal profitable
activities or product-market opportunities, managers may prefer to
aggregate operations rather than inform potential competitors
\autocite{hayes-lundholm-1996-segment,harris-1998-association,botosan-stanford-2005-managers}.
Firms' objections to the proposed standard reflected the same concern:
more disaggregated segment information could reveal proprietary
information and weaken competitive advantages
\autocite{ettredge-etal-2002-competitive,zhou-2022-proprietary}.

SFAS 131 therefore provides a useful setting for examining how customer
disclosures affect competition in upstream supplier markets. By
revealing segment-level performance and business composition that were
previously aggregated for proprietary reasons, the mandate may reduce
the information advantage that incumbent suppliers acquire through their
customer relationships. Whether this additional information alters
competition among suppliers is an empirical question.

\subsection{Customer Disclosure and Supplier-Side
Competition}\label{customer-disclosure-and-supplier-side-competition}

An incumbent supplier can learn about a customer through repeated
transactions and private communication. For example, customers may share
demand forecasts or engage in joint product development that reveals
information about customer needs unavailable to firms outside the
relationship
\autocite{handfield-etal-1999-involving,cachon-fisher-2000-supply}. Even
when the customer does not share demand forecasts explicitly, the
incumbent supplier can infer changes in demand from order histories and
interactions with the customer's business units. This
relationship-generated knowledge helps the supplier plan production and
choose capacity before potential suppliers can observe the same
opportunity.

Expanded segment disclosures can erode the incumbent supplier's
relationship-based information advantage. Public data on the customer's
segment sales and growth allow potential suppliers to identify where
downstream demand is concentrated and which customer businesses may
justify entry or product repositioning. Evidence from other settings
shows that financial reporting can help outside firms identify
product-market opportunities and intensify competition in the disclosing
firm's market \autocite{yang-2019-real,glaeser-omartian-2022-public}. By
narrowing the information asymmetry between incumbent and potential
suppliers, customer disclosure may enable potential suppliers to
reposition toward the incumbent supplier's product space.

However, expanded disclosures need not alter the competitive environment
faced by incumbent suppliers. Analysts and market participants possessed
some segment information before SFAS 131
\autocite{berger-hann-2003-impact}, and public segment data may be too
coarse to guide a potential supplier's entry decision. Moreover,
incumbent suppliers may retain competitive advantages arising from
established customer relationships and capabilities that cannot be
replicated from public disclosures. Whether expanded customer disclosure
is sufficient to alter supplier-side competition is therefore an
empirical question. I state the hypothesis in null form:\\

\noindent \textbf{H1.} Expanded customer segment disclosures do not
affect the competitive environment faced by incumbent suppliers.\\

\subsection{Customer Disclosure and Suppliers' Internal Capital
Allocation}\label{customer-disclosure-and-suppliers-internal-capital-allocation}

For a multi-segment supplier, customer disclosures can affect how the
supplier allocates capital across its segments. The disclosures can
reveal growth opportunities or competitive threats that differ across
the supplier's segments, thereby changing the relative attractiveness of
investment across segments. Because headquarters allocates scarce
capital across segments, additional investment in one segment generally
means fewer resources available for its sibling segments. Customer
disclosures can shape these allocation decisions through two channels:
by improving the supplier's information about downstream opportunities
(the ``information channel'') or by exposing these opportunities to
potential suppliers (the ``competitive-threat channel'').

Under the information channel, expanded customer disclosures reduce
uncertainty about downstream demand. Suppliers make
relationship-specific investments whose value depends partly on the
customer's future operations, but the customer is better informed about
those prospects. This information asymmetry can constrain otherwise
valuable supplier investment \autocite{williamson-1985-economic}.
Consistent with this reasoning, prior studies find that more informative
customer disclosures are associated with more efficient supplier
investment
\autocite{chiu-etal-2019-customers,chen-etal-2019-linguistic,chiu-etal-2022-how}.
Disaggregated segment reports can similarly help suppliers identify
which of their segments serve customer businesses with stronger
fundamentals. If this channel dominates, capital should become more
closely aligned with supplier-segment and customer-segment growth
opportunities. In particular, expanded disclosure should reduce
under-investment where uncertainty previously constrained investment.

Under the competitive-threat channel, the same disclosure informs
potential suppliers as well as the incumbent supplier. Once potential
suppliers can identify attractive downstream opportunities, the
incumbent supplier has an incentive to protect its customer
relationships and competitive position. Industrial organization theory
shows that committed capacity can deter entry by reducing the demand
available to potential competitors and their expected profits after
entry
\autocite{spence-1977-entry,dixit-1980-role,huisman-kort-2015-strategic}.
Survey evidence indicates that entry deterrence is an important response
to competitive threats \autocite{smiley-1988-empirical}, and empirical
evidence shows that threatened incumbents expand physical capacity and
that these commitments can reduce subsequent entry
\autocite{cookson-2018-anticipated}. In this setting, the supplier may
direct capital toward segments exposed by customer disclosure to expand
capacity and make entry less attractive to potential suppliers. If this
channel dominates, the supplier may invest beyond what the supplier
segment's growth opportunities would otherwise support.

These channels yield different directional predictions for internal
capital allocation. Better information should align investment more
closely with the growth opportunities, whereas heightened competitive
threats may lead suppliers to invest in exposed segments beyond what
those growth opportunities would support. Which effect dominates is ex
ante unclear. I therefore state the internal capital allocation
hypothesis in null form:\\

\noindent \textbf{H2.} Expanded customer segment disclosure does not
affect incumbent suppliers' internal capital allocation decisions.\\

\subsection{Distinguishing the Channels and Assessing an Alternative
Explanation}\label{distinguishing-the-channels-and-assessing-an-alternative-explanation}

The previous hypothesis, H2, is concerned with whether customer
disclosures affect suppliers' internal capital allocation decisions, but
evidence of such an effect would not identify the underlying channel.
The directional predictions developed above provide one way to
distinguish them. If disclosure alleviates information frictions that
previously constrained valuable investment, it should reduce
underinvestment. In contrast, the competitive-threat channel predicts
investment beyond the level supported by growth opportunities as
suppliers seek to defend exposed customer relationships.

The competitive-threat channel also predicts that the investment effect
should be stronger when supplier segments have more economic value at
risk from heightened competition. Relationships with larger customers
may be more economically valuable to supplier segments, strengthening
suppliers' incentives to protect those relationships when competitive
threats increase. Similarly, supplier segments operating in more
concentrated industries may have greater incumbent rents at risk when
competitive pressure increases. Accordingly, the competitive-threat
channel predicts larger increases in investment when affected supplier
segments are linked to larger customers or operate in more concentrated
industries. These predictions are less directly implied by the
information channel, under which investment adjustments depend primarily
on the extent to which disclosure resolves uncertainty about growth
opportunities.

Finally, the subsequent outcomes associated with these investments
provide an additional way to distinguish the two channels. If investment
reflects a response to competitive threats, suppliers may accept the
cost of additional capacity in exchange for preserving customer
relationships or market position. Accordingly, greater reallocation may
be associated with a larger subsequent customer base or greater market
share. At the same time, defensive capacity investment may reduce
subsequent profitability when the additional capacity exceeds that
supported by underlying growth opportunities. In contrast, if disclosure
alleviates underinvestment, the resulting investment should not predict
lower subsequent operating performance.

In addition to distinguishing between the information and
competitive-threat channels, I consider a separate mechanism through
which customer disclosure could affect supplier investment. Customer
disclosure may increase competition in the customer's own product
market, leading the customer to adjust its operations and demand for
inputs. The supplier may then expand capacity to accommodate changes in
customer demand rather than to counter competitive threats from
potential suppliers. This alternative implies that expanded disclosure
should increase competition faced by disclosing customers and that
supplier investment should increase more when linked customers
experience greater increases in competition. These predictions provide a
way to distinguish changes in customer-driven input demand from the
supplier-side competitive-threat channel.

\section{Data Collection and Sample Selection}\label{sec:data_sample}

\subsection{Data Collection}\label{sec:data_collection}

This section details the data collection process. To analyze internal
capital allocations, I construct a dataset that links supplier segments
to specific customer segments. This process requires three primary
sources: the Compustat Segment database for segment-level financials,
the Compustat WRDS Supply Chain with IDs (hereafter ``Supply Chain
database'') for customer identities, and the Bureau of Economic Analysis
(BEA) Input-Output data (hereafter ``BEA IO data'') to establish the
economic linkages between supplier segments and customer segments.

The Compustat Segment database provides segment-level sales, capital
expenditures, and assets, together with SIC industry codes for each
segment. A practical challenge is that the database does not
consistently retain the restated segment information needed for firms
that revised their reporting structure after SFAS 131 (``change
firms'').\footnote{In \textcite{cho-2015-segment}, a change firm is one
  that modifies segment disclosure after SFAS 131 and whose restated
  post-SFAS 131 segments reveal operations in industries that were not
  previously disclosed. For the customer-disclosure tests in this paper,
  I additionally require the customer's number of reported segments to
  increase after adoption so that the disclosure change reflects a
  measurable expansion in transparency.} Without the restated
disclosures, a change firm can appear to have a different segment
structure before and after adoption, which prevents a consistent measure
of internal capital allocation over time. Following
\textcite{cho-2015-segment}, I therefore hand-collect the pre-SFAS 131
restated segment data for these firms from 10-K filings on EDGAR. This
procedure keeps segment definitions comparable across the pre- and
post-adoption periods.

Next, I extract supplier-customer relationships from the Supply Chain
database constructed by \textcite{cohen-frazzini-2008-economic} and
\textcite{cen-etal-2017-customer}. This database identifies major
customers, typically those accounting for at least 10\% of a supplier's
sales. However, as noted by \textcite{ellis-etal-2012-proprietary}, this
database often contains significant gaps in supplier-customer linkages
due to non-disclosure. To address these gaps, I manually review
suppliers' 10-K filings via the SEC's EDGAR database and establish
additional customer links omitted from the Supply Chain database.
Specifically, for suppliers that disclose customer identities in some
years but omit them in others, I compare reported revenue percentages
across years to consistently identify major customers throughout the
sample period (see \cref{sec:example_hand_customer} for a detailed
example of this procedure).

To link a supplier's segments to a customer's segments, I utilize the
BEA IO data, specifically the Direct Domestic Requirements table
following \textcite{carter-etal-2021-effect}. This table outlines the
input-output relationship among U.S. industries starting in
1997.\footnote{The requirements tables are available at the Bureau of
  Economic Analysis website:
  \url{https://www.bea.gov/industry/input-output-accounts-data}.} I
convert the BEA's industry ``IO codes'' to NAICS codes and SIC codes to
align with the segment SIC industry codes used in the Compustat Segment
database.

The BEA IO data indicate the amount of input required from each supplier
industry to produce one dollar of output in a given customer industry.
Using this information, I rank supplier segments for a given customer
segment based on the strength of the input-output relationship. The
supplier segment with the strongest input-output linkage to the
customer's segment is identified as the primary match (see
\cref{sec:example_seg_seg_matching} for an example).

\subsection{Sample Selection}\label{sec:sample_selection}

\cref{tbl:s19_01_sample_selection} presents the sample selection
procedure. I begin with 47,010 firm-years (12,434 firms) represented in
Compustat segment data between fiscal years 1996 and 2000. Requiring a
linked major customer, based on the Supply Chain database augmented with
hand-collected links, removes 37,948 supplier-years, leaving 9,062
supplier-years from 3,499 suppliers linked to 2,122 unique major
customers. Customer counts throughout the table refer to unique linked
major customers.

Following \textcite{berger-hann-2003-impact} and
\textcite{cho-2015-segment}, I next exclude suppliers whose
segment-structure changes reflect real operating changes, such as
mergers, acquisitions, or restructurings.\footnote{\textcite{cho-2015-segment}
  implements the \textcite{berger-hann-2003-impact} algorithm and
  excludes change firms when the difference in aggregate segment
  revenues or earnings between the restated SFAS 131 report and the
  corresponding historical pre-SFAS 131 report exceeds 1\% of the
  restated total.} This screen removes 3,582 supplier-years and leaves
5,480 supplier-years from 1,803 suppliers. Excluding suppliers in
financial (SIC 6000--6999) and regulated (SIC 4900--4949) industries
removes another 72 supplier-years, leaving 5,408 supplier-years from
1,777 suppliers. Requiring complete segment assets, capital
expenditures, and sales then removes 4,101 supplier-years, leaving 1,307
supplier-years from 444 suppliers.

I then require at least one linked major customer in the SFAS 131
classification sample from \textcite{cho-2015-segment}. This restriction
removes 166 supplier-years and leaves 1,141 supplier-years from 386
suppliers. Requiring the treatment variables removes 239 supplier-years
and leaves 902 supplier-years from 318 suppliers. Finally, requiring
complete controls removes five supplier-years. The firm-level regression
sample therefore contains 897 supplier-years from 316 suppliers linked
to 386 unique major customers.

The segment-level investment tests begin with the Compustat Segment
database. I remove segments that appear only in the pre-SFAS 131 period
or only in the post-SFAS 131 period so that changes in the estimates are
not driven by newly created or discontinued segments. I also exclude
segments with no substantive operations, such as unallocated or
reconciling segments, and I drop geographic segments to focus on
business segments. The final segment sample consists of 1,071 unique
supplier-segment-years with 263 unique suppliers and 4,445
supplier-segment-customer-segment-year pairs.

\subsection{Treatment Intensity and Descriptive
Statistics}\label{treatment-intensity-and-descriptive-statistics}

The main analysis uses a continuous measure of treatment intensity,
\(\% Treat\), defined as the proportion of a supplier's major customers
that expanded their segment disclosures following SFAS 131. \(Post\) is
an indicator equal to one for fiscal years following the implementation
of SFAS 131 and zero otherwise. For descriptive purposes only, I
partition suppliers into high- and low-\(\% Treat\) groups based on the
sample median of \(\% Treat\) and report descriptive statistics
separately for the pre- and post-SFAS 131 periods in
\cref{tbl:desc_firm}. The empirical analyses retain the continuous
measure of treatment intensity rather than this median-based partition.

Because the sample median of \(\% Treat\) is 1, the high-\(\% Treat\)
group consists of suppliers whose major customers all expanded their
segment disclosures following SFAS 131. Among suppliers in the
low-\(\% Treat\) group, approximately 25\% of major customers expanded
their segment disclosures.

Overall, firm characteristics are broadly comparable across the four
subsamples. Relative to suppliers in the low-\(\% Treat\) group,
suppliers in the high-\(\% Treat\) group are larger, have lower
market-to-book ratios, higher operating cash flows, are more likely to
invest in R\&D or intangible assets, have fewer reported segments, and
operate in less concentrated industries.

I conduct a parallel exercise for the segment-level descriptive
statistics in \cref{tbl:desc_seg}. At the segment level, treatment is
binary: I classify a segment as treated if its linked customer expanded
its segment disclosures following SFAS 131 and as a control segment
otherwise. I report descriptive statistics separately for treated and
control segments in the pre- and post-SFAS 131 periods. Overall, the
descriptive statistics across the four subsamples are broadly
comparable. Compared with the control segments, the treated segments are
larger in absolute terms, face similar industry-wide investment
opportunities, as proxied by industry Tobin's \(q\), exhibit higher
segment sales growth, and are linked to larger customers.

\section{Research Design}\label{sec:research_design}

The empirical analysis proceeds in three stages. First, I test whether
customer disclosures increase suppliers' exposure to competition.
Second, I examine whether those disclosures change how suppliers
allocate capital across their own segments. Third, I use segment-level
investment tests to distinguish whether the observed reallocation
reflects reduced under-investment arising from better information about
customer demand (the ``information channel'') or investment undertaken
to counter competitive threats in exposed segments beyond what growth
signals would predict (the ``competitive-threat channel'').

\subsection{Suppliers' Competitive
Environment}\label{suppliers-competitive-environment}

I measure the competitiveness of the supplier industry using two
text-based measures derived from the Text-based Network Industry
Classifications
\autocites[TNIC-3,][]{hoberg-phillips-2010-product,hoberg-phillips-2016-textbased}:
the number of rivals and textual similarity of product descriptions.
TNIC is well suited for this setting because customer disclosures may
invite repositioning by potential suppliers that are not yet captured by
standard industry definitions or concentration ratios. Text-based rivals
therefore provide a useful way to detect whether potential suppliers
move closer to an incumbent supplier's product space after the
customer's disclosure becomes more informative.

To quantify the extent to which a supplier is exposed to customer
disclosure expansion, I use a continuous treatment variable,
\(\% Treat\). It is defined as the ratio of the supplier's major
customers that expanded segment disclosure following SFAS 131 to the
supplier's total number of major customers. For example, if a supplier
has three major customers and one of them expands segment disclosure,
the supplier's \(\% Treat\) equals 1/3. This example is further
elaborated in \cref{sec:example_treat_post}.

To examine the effect of customer disclosures on suppliers' competitive
environment, I estimate the following difference-in-differences
specification:
\begin{equation}\protect\phantomsection\label{eq:supplier_competition_did}{
\begin{aligned}
Competition_{it}
={}& \beta \left(\% Treat_{it} \times Post_t\right) + \gamma Controls_{it} \\
&+ \mathit{Firm\ FE}_i + \mathit{Year\ FE}_t + \varepsilon_{it},
\end{aligned}
}\end{equation} where \(Competition\) is either \(N\text{ }Rivals\) or
\(Textual\text{ }Similarities\). \(N\text{ }Rivals\) is the number of
rivals defined by TNIC-3, and \(Textual\text{ }Similarities\) is the
overall similarity of product descriptions with those rivals
\autocite{hoberg-phillips-2010-product,hoberg-phillips-2016-textbased}.
\(Post\) is an indicator variable equal to one for the post-SFAS 131
period and zero otherwise. \(Controls\) is a vector of firm-level
characteristics defined in \cref{sec:var_def}. The baseline
specification includes supplier-firm and year fixed effects. I also
estimate a specification that replaces year fixed effects with two-digit
SIC industry-year fixed effects to absorb industry-specific shocks such
as merger waves or common changes in demand. A positive coefficient on
\(\% Treat \times Post\) indicates that suppliers exposed to more
disclosing customers face a more competitive product-market environment
after SFAS 131. Standard errors are clustered at the supplier-firm
level.

\subsection{Firm-level Internal Capital
Allocations}\label{sec:cho_int_eff}

I measure firm-level internal capital allocation using a benchmark
derived from \textcite{cho-2015-segment}. The idea is to compare each
segment's share of firm capital expenditures with the share that would
arise if the firm allocated capital across segments in proportion to
segment sales. The difference between those two shares captures how far
the firm departs from that benchmark:
\[CAPX\text{ }deviation_{ijt} = \frac{CAPX_{ijt}}{\sum_{j=1}^n CAPX_{ijt}} - \frac{Sale_{ijt}}{\sum_{j=1}^n Sale_{ijt}},\]
where \(CAPX\) represents segment-level capital expenditures, and
\(Sale\) is segment-level sales. \(i\), \(j\), and \(t\) denote a firm,
a segment, and a year, respectively, and \(n\) is the total number of
segments of a firm \(i\).

A sign is then assigned to that difference based on whether the segment
has stronger or weaker benchmark opportunities than its sibling
segments. Segment opportunities are proxied by the one-year-lagged
median Tobin's \(q\) of single-segment firms in the same SIC industry.
Using lagged \(q\) ensures that the opportunity benchmark is measured
before the capital allocation in year \(t\). When a segment with
stronger benchmark opportunities receives relatively more capital, the
signed measure increases; when a segment with weaker benchmark
opportunities receives relatively more capital, the signed measure
decreases. Formally, the Signed CAPX deviation is: \[\begin{aligned}
Signed\text{ }CAPX\text{ }deviation_{ijt} &= (+1) \cdot CAPX\text{ }deviation_{ijt} \text{ if } q_{ij,t-1} > \bar{q}_{i,t-1},\\
Signed\text{ }CAPX\text{ }deviation_{ijt} &= (-1) \cdot CAPX\text{ }deviation_{ijt} \text{ if } q_{ij,t-1} \leq \bar{q}_{i,t-1},
\end{aligned}\] where \(q_{ij,t-1}\) is lagged Tobin's \(q\) for segment
\(j\) of firm \(i\) and \(\bar{q}_{i,t-1}\) is the
segment-asset-weighted average lagged \(q\) of firm \(i\)'s other
segments. \(q_{ij,t-1}\) is proxied by the lagged median \(q\) of
single-segment firms within the same industry as segment \(j\).

Lastly, the segment-level signed values are aggregated using segment
assets as weights to get the \textcite{cho-2015-segment} firm-level
internal capital allocation measure: \[ \begin{aligned}\
Capital\text{ }Allocation\text{ }Efficiency_{it} &= \sum_{j=1}^n w_{ijt} \cdot Signed\text{ }CAPX\text{ }deviation_{ijt},\\
\text{where } w_{ijt} &= \frac{BA_{ijt}}{\sum_{j=1}^n BA_{ijt}}.
\end{aligned}\] \(BA\) represents the segment assets. \(i\), \(j\), and
\(t\) denote a firm, a segment, and a year, respectively, and \(n\) is
the total number of segments of a firm \(i\).

I next estimate the effect of customer disclosures on the supplier's
firm-level allocation measure:
\begin{equation}\protect\phantomsection\label{eq:firm_capital_allocation_did}{
\begin{aligned}
Capital\text{ }Allocation\text{ }Efficiency_{it}
={}& \beta \% Treat_{it} \times Post_t + \gamma Controls_{it} \\
&+ \mathit{Firm\ FE}_i + \mathit{Year\ FE}_t + \varepsilon_{it},
\end{aligned}
}\end{equation} where \(Post\) is an indicator variable equal to one in
the post-SFAS 131 period and zero otherwise. \(Controls\) contains
firm-level characteristics defined in \cref{sec:var_def}. The baseline
specification includes supplier-firm and year fixed effects. I also
estimate an alternative specification that replaces year fixed effects
with two-digit SIC industry-year fixed effects. In this paper's setting,
a negative coefficient on \(\% Treat \times Post\) indicates that
treated suppliers shift relatively more capital toward segments with
weaker benchmark growth opportunities than their sibling segments.
Standard errors are clustered at the supplier-firm level.

\subsection{Segment-level Investments}\label{sec:biddle_seg_eff}

To analyze segment-level investment decisions, I adapt the sales growth
model of investment \autocite{biddle-etal-2009-how} to the segment level
and augment it with the growth of the linked customer segment. The first
step estimates expected segment investment as a function of both signals
(sales growth of the supplier segment and that of the customer segment):
\[ SegInvest_{j, t+1} = \beta_0 + \beta_1 \Delta SegSales_{jt} + \beta_2 \Delta Customer\text{ }Sales_{kt} + \varepsilon_{jt}, \]
where \(SegInvest\) is a supplier segment's lead capital expenditures
(in year \(t+1\)) divided by the segment assets (in year \(t\)),
\(\Delta SegSales_{jt}\) indicates the sales growth (from year \(t-1\)
to year \(t\)) of supplier segment \(j\),
\(\Delta Customer\text{ }Sales_{kt}\) represents the sales growth of the
corresponding customer segment \(k\). \(i\), \(j\), \(k\), and \(t\)
denote a supplier, a supplier segment, a customer segment linked to the
supplier segment \(j\), and a year, respectively. This regression is run
by Fama-French 48 industry and year with at least 10 observations for a
given industry and year. The residual from the first-step regression is
defined as \emph{Residual\_InvDev}. Because these first-step regressions
serve only to construct residual measures, I do not draw inferences from
their coefficient estimates or standard errors.

I then convert the residuals from that model into one efficiency measure
and two deviation measures. \(InvEff\text{ }(Abs)\) is the absolute
value of \emph{Residual\_InvDev} multiplied by \ensuremath{-}1. Thus, higher values
imply investment closer to the benchmark implied by sales growth.
\emph{Positive Investment Residual} equals the residual when the
residual is positive and zero otherwise. \emph{Negative Investment
Residual} equals \((-1)\cdot Residual\_InvDev\) when the residual is
negative and zero otherwise. This decomposition is important for
distinguishing between channels: the competitive-threat channel predicts
that treated segments will invest more than growth signals warrant
(higher positive residuals), whereas the information channel predicts
that treated segments will reduce under-investment relative to growth
signals (lower negative residuals).

I proceed with the following segment-level difference-in-differences
specification:
\begin{equation}\protect\phantomsection\label{eq:segment_investment_did}{
\begin{aligned}
Investment\text{ }Efficiency/Deviation_{ijk,t+1}
={}& \beta \left(Treat_{ijk} \times Post_t\right) + \gamma Controls_{ijkt} \\
&+ \mathit{Supplier\!-\!Segment\ FE}_{ij} \\
&+ \mathit{Industry\!-\!Year\ FE}_{Industry(j)t} + \varepsilon_{ijkt},
\end{aligned}
}\end{equation} where
\(Investment\text{ }Efficiency/Deviation_{ijk,t+1}\) corresponds to one
of the investment efficiency (or investment deviation) measures defined
above. \(i\), \(j\), \(k\), and \(t\) denote a supplier, a supplier
segment, a customer segment linked to supplier segment \(j\), and a
year, respectively. \(Treat\) equals one if the linked customer segment
belongs to a customer that expanded segment disclosure following SFAS
131, and zero otherwise. \(Post\) equals one for the post-SFAS 131
period, and zero otherwise. \(Controls\) include the first-step
independent variables, segment characteristics following
\textcite{benz-etal-2023-picking}, and matched customer-segment
characteristics, all defined in \cref{sec:var_def}. The specification
includes supplier-segment and industry-year fixed effects, where
industries are defined at the Fama-French 48 level, consistent with the
first-step grouping. I follow \textcite{chen-etal-2018-incorrect} and
\textcite{chen-etal-2022-use} in using this two-step design because the
first-step model is estimated within industry-year groups.\footnote{While
  \textcite{chen-etal-2018-incorrect} and \textcite{chen-etal-2022-use}
  recommend a single-regression approach, they also suggest a two-step
  approach with all the first-step independent variables in the
  second-step regression. As such, I employ a two-step approach in this
  study. This choice is motivated by the need to distinguish between
  positive and negative investment residuals at the segment level, which
  is crucial for identifying the source of changes in firm-level
  internal capital allocations. The two-step approach allows for a more
  nuanced analysis of investment deviation measures, enabling a clearer
  interpretation of how customer disclosures affect suppliers'
  segment-level investment decisions.} Standard errors are clustered at
the firm-segment level.

The coefficient of interest is \(\beta\). A significantly positive
coefficient in the \emph{Positive Investment Residual} regressions
indicates that treated segments invest more than the sales-growth
benchmark would predict after customer disclosure expands, consistent
with the competitive-threat channel. A significantly negative
coefficient in the \emph{Negative Investment Residual} regressions would
instead indicate that treated segments reduce underinvestment relative
to the benchmark, consistent with the information channel.

\subsection{Supplier Outcomes Associated with Capital
Reallocation}\label{sec:reallocation_outcomes}

I next examine whether larger capital reallocations are associated with
suppliers' subsequent market position and customer relationships. I
define \emph{Capital Reallocation Magnitude} as \emph{Capital Allocation
Efficiency} multiplied by \(-1\), so higher values indicate larger
deviations from the benchmark. For sales-based market share, I estimate:
\begin{equation}\protect\phantomsection\label{eq:market_share_reallocation}{
\begin{aligned}
Market\text{ }Share_{i,\ t+1\text{ or }t+2}
={}& \beta_1 Post_t \times Capital\text{ }Reallocation\text{ }Magnitude_{it} \\
&+ \beta_2 Capital\text{ }Reallocation\text{ }Magnitude_{it} + \gamma Controls_{it} \\
&+ \mathit{Firm\ FE}_i + \mathit{Year\ FE}_t + \varepsilon_{it},
\end{aligned}
}\end{equation} where \(Market\text{ }Share\) is firm sales divided by
total sales in the firm's three-digit SIC industry. The coefficient
\(\beta_1\) captures whether the association between reallocation
magnitude and subsequent market share differs in the post-SFAS 131
period.

For the number of customers, I estimate the following OLS specification:
\begin{equation}\protect\phantomsection\label{eq:customer_count_reallocation}{
\begin{aligned}
N\text{ }Customers_{i,\ t+1\text{ or }t+2}
={}& \beta_1 Post_t \times Reallocation\text{ }Magnitude_{it} \\
&+ \beta_2 Reallocation\text{ }Magnitude_{it} + \gamma Controls_{it} \\
&+ \mathit{Firm\ FE}_i + \mathit{Year\ FE}_t + \varepsilon_{it},
\end{aligned}
}\end{equation} where \(N\text{ }Customers\) is the number of customers
in year \(t+1\) or \(t+2\). For both the market-share and customer-count
specifications, standard errors are clustered at the supplier-firm
level.

Finally, I examine whether supplier segments with larger positive
investment residuals subsequently experience lower profitability, as
would be expected if investment is undertaken to defend the supplier's
competitive position:
\begin{equation}\protect\phantomsection\label{eq:segment_roa_reallocation}{
\begin{aligned}
Segment\text{ }ROA_{ij,\ t+1\text{,\ }t+2\text{ or }t+3}
={}& \beta_1 Post_t \times Positive\text{ }Investment\text{ }Residual_{ijt} \\
&+ \beta_2 Positive\text{ }Investment\text{ }Residual_{ijt} \\
&+ \gamma Controls_{ijt} + \mathit{Firm\ FE}_i + \mathit{Year\ FE}_t + \varepsilon_{ijt},
\end{aligned}
}\end{equation} where \(Segment\text{ }ROA\) is segment profit scaled by
lagged segment assets in the subsequent year \(t+1\), \(t+2\), or
\(t+3\). A negative \(\beta_1\) is more consistent with investment
undertaken in response to competitive threats than with improved
allocation driven by new information. Standard errors are clustered at
the supplier-firm level.

\subsection{Customer-side Competition
Alternative}\label{sec:customer_competition_alternative}

An alternative explanation is that suppliers are not responding to
intensified competition in their own product markets, but instead
expanding capacity in response to heightened competition in their
customers' product markets following greater segment disclosures. I
examine this possibility in two ways.

First, I move the unit of analysis to the customer firm and test whether
a customer's own expanded segment disclosure changes its product-market
environment:
\begin{equation}\protect\phantomsection\label{eq:customer_competition_did}{
\begin{aligned}
CustomerCompetition_{ct}
={}& \beta \left(Treat_c \times Post_t\right) + \gamma Controls_{ct} \\
&+ \mathit{Customer\ Firm\ FE}_c + \mathit{Year\ FE}_t + \varepsilon_{ct},
\end{aligned}
}\end{equation} where \(CustomerCompetition\) is either the customer's
TNIC-3 rival count or textual similarity to its TNIC-3 rivals.
\(Treat_c\) equals one for customers that expanded segment disclosure
following SFAS 131, and zero otherwise. \(Post_t\) equals one in the
post-SFAS 131 period, and zero otherwise. I estimate specifications with
customer-firm and year fixed effects, and specifications that replace
year fixed effects with two-digit SIC industry-year fixed effects.
Standard errors are clustered at the customer-firm level. If the
customer's-competitive-environment alternative is correct, treated
customers should exhibit an increase in their own product-market
competition after their disclosures expand.

Second, I test whether the supplier's internal-capital-allocation effect
is stronger when linked customers experience larger post-disclosure
increases in rival counts. For each supplier, I compute the average
change in TNIC-3 rival counts among its linked customers, comparing the
post-SFAS 131 period with the pre-SFAS 131 period. I use both a
standardized continuous measure of that change and tercile indicators
for middle and high customer-rival-count changes. I then estimate:
\begin{equation}\protect\phantomsection\label{eq:customer_competition_reallocation}{
\begin{aligned}
Capital\text{ }Allocation\text{ }Efficiency_{it}
={}& \beta_1 \left(\% Treat_{it} \times Post_t\right) \\
&+ \beta_2 \left(\% Treat_{it} \times Post_t
  \times \Delta CustomerCompetition_i\right) \\
&+ \gamma Controls_{it} + \mathit{Firm\ FE}_i + \mathit{Year\ FE}_t + \varepsilon_{it}.
\end{aligned}
}\end{equation} The coefficient \(\beta_2\) captures whether suppliers
reallocate more strongly when their customers experience larger
increases in competition. The customer's-competitive-environment
alternative predicts that the supplier response should be concentrated
among suppliers linked to customers with larger increases in their own
rival counts. Standard errors are clustered at the supplier-firm level.

\subsection{Parallel Trends Tests}\label{sec:parallel_trends_design}

To assess the parallel trends assumption for the two main investment
outcomes, I estimate event-study versions of
\cref{eq:firm_capital_allocation_did} and
\cref{eq:segment_investment_did}. The firm-level specification is:
\begin{equation}\protect\phantomsection\label{eq:firm_capital_allocation_event_study}{
\begin{aligned}
Capital\text{ }Allocation\text{ }Efficiency_{it}
={}& \sum_{\tau \in \{1997,1999,2000\}} \beta_\tau
  \left(\% Treat_{it} \times Year_\tau\right) \\
&+ \gamma Controls_{it} + \mathit{Firm\ FE}_i + \mathit{Year\ FE}_t + \varepsilon_{it}.
\end{aligned}
}\end{equation}

The corresponding supplier-segment specification is:
\begin{equation}\protect\phantomsection\label{eq:segment_investment_event_study}{
\begin{aligned}
Positive\text{ }Investment\text{ }Residual_{ijk,t+1}
={}& \sum_{\tau \in \{1997,1999,2000\}} \beta_\tau
  \left(Treat_{ijk} \times Year_\tau\right) \\
&+ \gamma Controls_{ijkt} + \mathit{Supplier\!-\!Segment\ FE}_{ij} \\
&+ \mathit{Industry\!-\!Year\ FE}_{Industry(j)t} + \varepsilon_{ijkt}.
\end{aligned}
}\end{equation} where \(Year_\tau\) is an indicator for year \(\tau\).
The year immediately before SFAS 131 adoption, \(Year_{1998}\), is
omitted for comparison. The firm-level specification clusters standard
errors by firm, and the supplier-segment specification clusters them by
firm-segment.

\section{Empirical Results}\label{sec:emp_results}

\subsection{Customer Disclosures and Supplier-side
Competition}\label{customer-disclosures-and-supplier-side-competition}

\cref{tbl:n_rivals_competition} presents estimates of
\cref{eq:supplier_competition_did}. Columns (1) -- (4) indicate that the
treated suppliers experience a significant increase in the number of
rivals in the post-SFAS 131 period compared with the control group. The
findings remain consistent across different fixed effects specifications
of firm and year fixed effects, and firm and industry-year fixed
effects. With a pre-period average of 34 rivals among treated suppliers,
the observed increase of 7 rivals (approximately 20\%) is economically
meaningful. This result suggests that expanded customer disclosures
alter the competitive environment faced by suppliers, providing a
mechanism through which customer disclosures affect suppliers' internal
capital allocation decisions.

Columns (5) -- (8) demonstrate that the product descriptions of the
treated suppliers become more similar to those of their rivals. The
coefficients are statistically significant and consistent across
different fixed effects structures. The increase in the textual
similarities is approximately 10\% relative to the pre-period average.
The rise in textual similarity suggests that potential suppliers
reposition their product offerings toward the incumbent supplier's
product space.

\subsection{Firm-level Internal Capital
Allocation}\label{firm-level-internal-capital-allocation}

\cref{tbl:main_results_firm} presents estimates of
\cref{eq:firm_capital_allocation_did}. Columns (1) -- (4) in
\cref{tbl:main_results_firm} present the effect of customer disclosures
on the supplier's internal capital allocation decisions. The significant
and negative coefficient for \(\% Treat \times Post\) suggests that the
allocation of treated suppliers has deviated from the benchmark
allocation implied by Tobin's \(q\) following the expanded customer
disclosures. The findings remain consistent across different fixed
effects specifications of firm and year fixed effects, and firm and
industry-year fixed effects. This result indicates that treated
suppliers direct their capital allocation to segments with lower growth
opportunities relative to Tobin's \(q\) by about 6\% of total capital
expenditure compared with control suppliers post-SFAS 131.

\subsection{Firm-level Internal Capital Allocation, Market Share, and
Customer
Base}\label{firm-level-internal-capital-allocation-market-share-and-customer-base}

I examine whether suppliers' capital reallocation in response to
competitive threats is associated with maintaining their customer
relationships. I explore the relationship between capital reallocation
magnitude and two measures of customer-market outcomes: sales-based
market share and the number of customers, using the specifications in
\cref{eq:market_share_reallocation} and
\cref{eq:customer_count_reallocation}, respectively.

Columns (1) and (2) of \cref{tbl:conseq_01_firm_customer} report
estimates of \cref{eq:market_share_reallocation} for sales-based market
share in the subsequent two years. A significantly positive coefficient
\(\beta_1\) for
\(Post_t \times Capital\text{ }Reallocation\text{ }Magnitude_{it}\) in
Column (2) suggests that suppliers with larger capital reallocations
tend to secure a larger market share in the post-SFAS 131 period. The
results are robust to alternative specifications of the market share,
such as the weighted average of segment-level market share and weighted
average of segment-level market share from segments that experience
increased customer disclosures.

Column (4) of \cref{tbl:conseq_01_firm_customer} reports estimates of
the OLS specification in \cref{eq:customer_count_reallocation}. The
significantly positive coefficient \(\beta_1\) on
\(Post_t \times Reallocation\text{ }Magnitude_{it}\) indicates that the
association between reallocation magnitude and the subsequent number of
customers is more positive in the post-SFAS 131 period, consistent with
suppliers maintaining or expanding their customer base.

\subsection{Segment-level Investment}\label{segment-level-investment}

Next, I turn to segment-level analysis. \cref{tbl:main_results_seg}
presents estimates of \cref{eq:segment_investment_did}, where
\(Investment\text{ }Efficiency/Deviation_{ijk,t+1}\) represents one of
the following investment efficiency (or investment deviation) measures:
\emph{InvEff (Abs)}, \emph{Positive Investment Residual}, and
\emph{Negative Investment Residual}. Supplier-segment fixed effects and
industry-year fixed effects are included following
\textcite{chen-etal-2018-incorrect} and \textcite{chen-etal-2022-use}
because the first-step regression is run by industry and year. Columns
(1) -- (2) in \cref{tbl:main_results_seg} indicate that supplier
segments exhibit lower investment efficiency relative to the
sales-growth benchmark after a customer's segment disclosure increases.
The effect is mainly driven by the \emph{Positive Investment Residual}
measure, which captures investment beyond what the sales growth model
would predict. Given that the pre-period mean of \emph{Positive
Investment Residual} is 0.031, the coefficient of 0.0093 is economically
significant, accounting for 30\% of the pre-period mean.

\subsection{Competitive-Threat
Channel}\label{competitive-threat-channel}

To strengthen the inferences and pin down the channel, I conduct two
cross-sectional tests based on the customer size and the level of
industry concentration among supplier segments. I expect the effect on
the \emph{Positive Investment Residual} measure to be stronger for the
subsample with larger customer size. The reason is that larger customers
are typically more crucial to suppliers and represent a significant
portion of their revenues. Consequently, suppliers may allocate more
resources toward sustaining relationships with larger customers compared
to those with smaller customers.

To examine this prediction, I augment \cref{eq:segment_investment_did}
by interacting \(Treat \times Post\) with \emph{High Customer Size}
while retaining the baseline \(Treat \times Post\) term and controls.
\emph{High Customer Size} equals one if customer size is above the
median and zero otherwise. The cross-sectional split of the sample is
based on characteristics during the pre-period. The definitions for all
other variables remain consistent with prior descriptions.

Consistent with the prediction, I find a significantly negative
coefficient on the triple interaction for the \emph{InvEff (Abs)}
measure in column (1) of \cref{tbl:xsect_seg_01_csize}. The coefficient
for the \emph{Positive Investment Residual} measure in column (2) is
positive and statistically significant. Thus, supplier segments linked
to larger customers exhibit greater absolute deviations from predicted
investment following SFAS 131 adoption than those associated with
smaller customers, and the change is driven by investments beyond what
growth signals would predict.

Next, I investigate whether the effect is more pronounced in supplier
segments operating in highly concentrated industries (in the
pre-period). In these industries, entry by potential suppliers poses a
greater competitive threat because it can erode incumbents' market
power. I therefore expect the effect to be stronger in industries with
higher pre-period industry concentration.

To evaluate this prediction, I augment \cref{eq:segment_investment_did}
by interacting \(Treat \times Post\) with \emph{High Supp Seg HHI} while
retaining the baseline \(Treat \times Post\) term and controls.
\emph{High Supp Seg HHI} equals one for supplier segments operating in
three-digit SIC industries with above-median concentration and zero
otherwise. The cross-sectional split of the sample is based on
characteristics during the pre-period.

Consistent with the prediction, I find a significantly negative
coefficient on the triple interaction for the \emph{InvEff (Abs)}
measure and a significantly positive coefficient for the \emph{Positive
Investment Residual} measure in columns (1) -- (2) of
\cref{tbl:xsect_seg_02_hhi}. This result indicates that supplier
segments in highly concentrated industries invest significantly more
following SFAS 131, compared with those in less concentrated industries.
This finding is consistent with suppliers reallocating capital toward
affected segments to protect their competitive position.

To further rule out the alternative explanation that increased
investments result from the mitigation of under-investment, I
investigate whether supplier segments with larger positive investment
residuals experience a decline in profitability.
\cref{tbl:conseq_02_seg_roa} reports estimates of
\cref{eq:segment_roa_reallocation}. Significant and negative
coefficients of \(\beta_1\) for
\(Post_t \times Positive\text{ }Investment\text{ }Residual_{ijt}\) in
Columns (2) and (3) imply that those supplier segments exhibit a decline
in segment-level profitability following larger positive investment
residuals in the post period. This evidence is more consistent with
reallocation undertaken in response to competitive threats than with
improved allocation driven by new information.

\subsection{Relationship between Firm-level Internal Capital Allocation
Efficiency Measure and Segment-level Investment Efficiency
Measure}\label{relationship-between-firm-level-internal-capital-allocation-efficiency-measure-and-segment-level-investment-efficiency-measure}

Throughout the paper, I use two measures of investment efficiency: the
firm-level internal capital allocation efficiency measure derived from
Tobin's \(q\) and the segment-level investment efficiency measure based
on the sales growth model. To establish that the two measures are
related and represent similar constructs of investment efficiency, I
analyze the relationship between the segment-level signed CAPX deviation
(described in \cref{sec:cho_int_eff}) and the segment-level investment
efficiency (defined in \cref{sec:biddle_seg_eff}) using a binscatter
plot \autocite[binned scatterplot,][]{cattaneo-etal-2024-binscatter}.

A binscatter plot is a visualization technique used to simplify and
clarify the relationship between two variables by partitioning the data
into bins and displaying the average outcome for each bin. Compared to
traditional scatter plots, this technique makes it easier to identify
the relationship. Binscatter plots also allow for the inclusion of
additional covariates, providing a clearer understanding of the
conditional mean function. \textcite{cattaneo-etal-2024-binscatter} also
propose the optimal number of bins and the optimal bin sizes,
considering the trade-offs between variance and bias. I adopt their
recommended approach in determining the number of bins and their sizes.

\cref{fig:binscatter_s18_08} depicts the conditional mean function of
the segment-level signed CAPX deviation measure in relation to the
segment-level investment efficiency measure, controlling for the
segment-level characteristics. The segment-level signed CAPX deviation
measure is increasing with the segment-level investment efficiency
measure. This figure illustrates that the two measures of investment
efficiency adopted in this study are positively associated with each
other.

\subsection{Additional Analysis}\label{additional-analysis}

To examine whether suppliers respond to the competitive threats through
margins other than internal capital reallocation, I test whether treated
suppliers increase advertising spending. The survey conducted by
\textcite{smiley-1988-empirical} identifies higher advertising spending
as the most widely used countermeasure. Because advertising expense data
are largely missing in Compustat \autocite{liang-2024-advertising}, I
supplement the analysis with SG\&A expenses. \cref{tbl:addl_adv_sga}
shows that treated suppliers increase both advertising and SG\&A
spending, consistent with suppliers responding to competitive threats
through additional expenditures.

An alternative explanation for my findings is that mandated segment
disclosures intensify the competitive environment of customer companies.
Consequently, supplier companies may be increasing production capacity
to meet the demands of these customers. In this explanation, supplier
companies would be allocating resources toward segments linked to
customers facing heightened competition, rather than allocating
resources in response to competition from potential suppliers.

I examine this possibility using the two tests described in
\cref{sec:customer_competition_alternative}. First, I test directly
whether customers that expand segment disclosures experience an increase
in their own product-market competition.
\cref{tbl:ia_customer_competition} reports estimates of
\cref{eq:customer_competition_did}, in which the dependent variables are
TNIC-3 rival counts and TNIC-3 textual similarity. Across specifications
with customer-firm and year fixed effects, and with customer-firm and
industry-year fixed effects, the coefficients on \(Treat \times Post\)
are statistically insignificant. Thus, there is no evidence that treated
customers themselves face an increase in product-market competition
after expanding segment disclosure.

Second, I test whether supplier reallocation is stronger when linked
customers experience larger increases in their own rival counts.
\cref{tbl:ia_customer_competition_reallocation} reports estimates of
\cref{eq:customer_competition_reallocation} using customer rival-count
changes measured continuously and by tercile. The interaction with the
standardized continuous change is close to zero, and the middle- and
high-tercile interactions are also statistically insignificant. At the
same time, the baseline \(Treat \times Post\) coefficient remains
negative in all specifications and statistically significant in four of
the five columns. These results indicate that the reallocation in
suppliers' investment is not concentrated among suppliers serving
customers whose own competitive environment changed most. This evidence
mitigates the concern that the reallocation is driven by their
customers' heightened competitive environment.

\subsection{Parallel Trends Assumption}\label{sec:parallel_trends}

An identifying assumption of the difference-in-differences (DiD)
research design is that the treatment and control groups would have
followed parallel outcome trends absent treatment. To validate this
assumption, \cref{fig:parallel_s14} displays the average firm-level
internal capital allocation efficiency for suppliers in the high- and
low-\(\% Treat\) groups across the years around the adoption of SFAS
131. 1997 and 1998 on the \(x\)-axis denote the pre-period, and 1999 and
2000 denote the post-period. The high-\(\% Treat\) group includes
suppliers with \(\% Treat\) at or above the sample median. Because the
median of \(\% Treat\) is 1, this group consists of suppliers whose
major customers all expanded their segment disclosures. The
low-\(\% Treat\) group includes suppliers with \(\% Treat\) below 1.

\cref{fig:parallel_s16} shows the equivalent exercise for the
segment-level mean of the \emph{Positive Investment Residual} variable.
The treatment group consists of supplier segments whose linked customer
expanded its segment disclosures. The control group consists of supplier
segments whose linked customer did not expand its segment disclosures.

In both figures, the mean values of the investment measures are
comparable in the pre-period. In the post-period,
\cref{fig:parallel_s14} shows that the firm-level internal capital
allocation efficiency measure declines for suppliers in the
high-\(\% Treat\) group relative to suppliers in the low-\(\% Treat\)
group. \cref{fig:parallel_s16} shows that the \emph{Positive Investment
Residual} increases for treated segments relative to control segments
(i.e., treated segments invest more than growth signals would predict).

In addition, I decompose the coefficient estimates of the firm-level
results in \cref{tbl:main_results_firm}. \cref{fig:coef_s14_93} plots
the \(\beta_\tau\) coefficients from
\cref{eq:firm_capital_allocation_event_study}. The figure indicates no
differential trend in firm-level internal capital allocation efficiency
before SFAS 131 adoption.

In a parallel analysis of the segment-level \emph{Positive Investment
Residual} measure, I decompose the coefficient estimates of the
segment-level results in \cref{tbl:main_results_seg}.
\cref{fig:coef_s19_03} plots the \(\beta_\tau\) coefficients from
\cref{eq:segment_investment_event_study}. The figure likewise indicates
no differential trend in positive investment residuals before SFAS 131
adoption.

A limitation of this study's parallel trends assumption test is the
restricted pre-period, which cannot extend beyond two years prior to the
adoption of SFAS 131. This constraint is mainly due to data
availability: most firms subject to SFAS 131 have restated their
pre-period segment data for a maximum of two years (represented by 1997
and 1998 in the figures).

\section{Conclusion}\label{sec:conclusion}

This paper examines whether and how customer disclosures affect
suppliers' internal capital allocation decisions. Using SFAS 131 as a
plausibly exogenous increase in customer segment disclosure, I find
evidence consistent with a competitive-threat channel rather than an
information channel. Following expanded customer disclosures, affected
suppliers face greater competition and reallocate capital toward
segments that appear weaker based on growth opportunities. Segment-level
analyses show that this reallocation reflects investment beyond that
predicted by segment growth signals. Importantly, these allocations have
strategic benefits but economic costs: suppliers that deviate more from
growth-based allocations subsequently preserve market share and expand
their customer base, while segments undertaking greater investment
relative to growth signals experience lower profitability. Taken
together, these findings suggest that customer disclosures affect
suppliers' internal capital allocation decisions through the
competitive-threat channel.

This study contributes to the disclosure and internal capital markets
literatures. I show that disclosures can shape how economically linked
firms allocate resources within their organizational boundaries. While
prior disclosure research documents spillover effects on peer firms'
aggregate investment decisions, this study shows that customer
disclosures affect the allocation of internal capital across supplier
segments. The findings also identify changes in the external disclosure
and competitive environment as a determinant of internal capital
allocation decisions. More broadly, the results highlight an important
consequence of greater transparency: by making customer information
available to both incumbent and potential suppliers, disclosures can
change competitive environments and induce strategic investment
decisions.

It is important to acknowledge that the generalizability of this study's
findings may be limited. The sample is restricted to suppliers with
major customers. Suppliers with less dependence on any single customer
might respond differently to customers' disclosures. Nevertheless, the
findings show that understanding the spillover effects of disclosure
requires considering both aggregate investment decisions and internal
capital allocation decisions.

\newpage

\singlespacing
\printbibliography

\newpage

% \appendix
\begin{appendices}
\renewcommand{\appendixname}{Appendix}
\renewcommand{\thesection}{\Alph{section}}

% Remove "Subsection" from subsections
\titleformat{\subsection}
  {\normalfont\large\bfseries}{\thesubsection}{1em}{}

\onehalfspacing

% \section{Appendix}
% var def
\section{Variable definitions}\label{sec:var_def}

\subsection{Firm-level variables}\label{firm-level-variables}

\subsubsection{Investment Efficiency
measures}\label{investment-efficiency-measures}

\begingroup
\setlength{\parindent}{1.5em}

\noindent \indent \emph{CAPX deviation}: the difference between the
ratio of segment CAPX to firm CAPX and the ratio of segment sales to
firm sales, i.e.,
\(\frac{\text{segment capital expenditure}}{\text{firm capital expenditure}} - \frac{\text{segment sales}}{\text{firm sales}}\).
\endgroup

\emph{Signed CAPX deviation}: the CAPX deviation if the segment's
one-year-lagged \(q\) (\emph{segment q}) exceeds the asset-weighted
average lagged \(q\) of other segments and
\((-1)\times CAPX\text{ }deviation\) otherwise.

\emph{Capital Allocation Efficiency}: the segment asset-weighted average
of the segment-level Signed CAPX deviation as illustrated in
\textcite{cho-2015-segment}. A higher value of \emph{Capital Allocation
Efficiency} indicates greater internal capital allocation efficiency.

\emph{Capital Reallocation Magnitude}: the \emph{Capital Allocation
Efficiency} measure, multiplied by -1. A higher value of \emph{Capital
Reallocation Magnitude} indicates a larger deviation from the Tobin's
\(q\) benchmark.

\subsubsection{Other variables}\label{other-variables}

\begingroup
\setlength{\parindent}{1.5em}

\noindent \indent \emph{Segment q}: the one-year-lagged median \(q\) of
single-segment firms operating in the same industry. \endgroup

\emph{\% Treat}: the proportion of a supplier's major customers that
expanded their segment disclosures following SFAS 131.

\emph{Post}: equal to one for fiscal years classified as post-SFAS 131
and zero for pre-period fiscal years.

\emph{N Rivals (Hoberg-Phillips)}: the number of rival companies based
on the Text-based Network Industry Classifications (TNIC-3). See
\textcite{hoberg-phillips-2010-product} and
\textcite{hoberg-phillips-2016-textbased} for more detail.

\emph{Textual Similarities (Hoberg-Phillips)}: the overall similarity of
product descriptions between a firm and all other firms in its product
market space. This measure is calculated by summing the pairwise
similarity scores between a focal firm and all other rival companies
based on the TNIC-3. According to product differentiation theory, this
measure is negatively associated with pricing power. Higher textual
similarity indicates less product differentiation and potentially more
intense competition. See \textcite{hoberg-phillips-2010-product} and
\textcite{hoberg-phillips-2016-textbased} for more detail.

\emph{log(Total Assets)}: the natural logarithm of the supplier firm's
total assets.

\emph{Market-to-Book}: the ratio of the market value of equity to the
book value of equity.

\emph{Cashflow}: the operating cash flows scaled by the lagged total
assets.

\emph{CapEx}: the capital expenditures scaled by the net PP\&E.

\emph{NonCapEx}: equal to one if a firm reports positive R\&D
expenditures or intangible assets and zero otherwise.

\emph{Tangibility}: the net PP\&E scaled by the total assets.

\emph{Cash}: cash and cash equivalents scaled by the total assets.

\emph{Leverage}: the ratio of total liabilities to the total assets.

\emph{Dividend}: equal to one if a firm reports positive dividends on
common stocks and zero otherwise.

\emph{External Financing}: the net external financing scaled by the
capital expenditures. The net external financing is computed as: sales
of common and preferred stock \(+\) long-term debt issuance \(-\)
purchases of common and preferred stock \(-\) long-term debt reduction
\(+\) any current debt changes \(-\) dividend.

\emph{N Segment}: the number of segments.

\emph{Speed of Profit Adjustment}: the asset-weighted average of the
profit adjustment speed in the industries where the segment operates.
The speed of profit adjustment is calculated from the following
regression:
\[ X_{ijt} = \beta_{0j} + \beta_{1j}D_n X_{ijt-1} + \beta_{2j}D_p X_{ijt-1} + \varepsilon_{ijt},\]
where \(X_{ijt}\) is the difference between a firm \(i\)'s ROA and the
industry-median ROA. \(D_n\) is an indicator variable equal to one if
\(X_{ijt-1}\) is negative and zero otherwise. \(D_p\) is an indicator
variable equal to one if \(X_{ijt-1}\) is positive and zero otherwise.
The regression is estimated in each industry for the past 20 years. The
estimated coefficient \(\beta_{2j}\) is the speed of profit adjustment
in industry \(j\). See \textcite{harris-1998-association} for the
detailed definition.

\emph{Concentration Ratio}: the asset-weighted average of the Herfindahl
index of the industries in which the segment operates.

\emph{Seg Earnings Persistence}: the asset-weighted average of the
persistence of abnormal earnings in the industries where the segment
operates. The persistence of abnormal earnings is calculated from the
regression similar to that of \emph{Speed of Profit Adjustment}. The
regression form is:
\[ X_{ijt} = \beta_{0j} + \beta_{1j}X_{ijt-1} + \varepsilon_{ijt},\]
where \(X_{ijt}\) is defined likewise. The regression is estimated in
each industry for the past 20 years. The estimated coefficient
\(\beta_{1j}\) is the segment earnings persistence in industry \(j\).
See \textcite{cho-2015-segment} for the detailed definition.

\emph{Seg Industry Diversity}: the ratio of the number of segments with
unique two-digit SIC codes to the total number of segments.

\emph{Market Share}: the ratio of a company's sales to the total sales
in the same three-digit SIC industry.

\emph{N Customers}: the number of customers linked to the supplier
company.

\emph{Advertising Expense}: the advertising expenses scaled by the total
assets.

\emph{SG\&A Expense}: the selling, general, and administrative expenses
scaled by the total sales.

\subsection{Segment-level variables}\label{segment-level-variables}

\subsubsection{Investment Efficiency
measures}\label{investment-efficiency-measures-1}

\begingroup
\setlength{\parindent}{1.5em}

\noindent \indent \emph{Residual\_InvDev}: the residual obtained from
the first-step regression of the segment investment (segment-level
capital expenditures scaled by lagged segment assets) on its
segment-level sales growth and the sales growth of the linked customer
segment by Fama-French 48 industry and year. \endgroup

\emph{InvEff (Abs)}: the absolute value of the \emph{Residual\_InvDev}.
This measure is multiplied by -1 to ease the interpretation, i.e., a
higher value of \emph{InvEff (Abs)} indicates greater investment
efficiency.

\emph{Positive Investment Residual} (or \emph{Positive Inv Resid}): a
truncated variable that assumes the value of \emph{Residual\_InvDev}
from the first-step regression when it is positive, and is set to zero
when it is negative. A higher value indicates that the segment invests
more than the sales growth model would predict.

\emph{Negative Investment Residual} (or \emph{Negative Inv Resid}): a
truncated variable that assumes the value of \emph{Residual\_InvDev}
from the first-step regression multiplied by -1 when it is negative, and
is set to zero when it is positive. A higher value indicates that the
segment invests less than the sales growth model would predict.

\subsubsection{Other variables}\label{other-variables-1}

\begingroup
\setlength{\parindent}{1.5em}

\noindent \indent \emph{Treat}: equal to one if the supplier segment is
linked to a customer that expanded its segment disclosures following
SFAS 131, and zero otherwise. \endgroup

\emph{Post}: equal to one for fiscal years classified as post-SFAS 131
and zero for pre-period fiscal years.

\emph{SegInvest}: lead capital expenditures of a supplier segment,
scaled by the segment assets.

\emph{Seg Sales Growth}: the percentage change in segment sales.

\emph{Customer Sales Growth}: the percentage change in the linked
customer's sales. If the segment of the customer is linked, then it is
the customer segment's sales growth. If the customer segment is not
linked, then it is imputed by the customer company's sales growth.

\emph{Seg Size}: the log value of the segment assets.

\emph{Seg Relative Size}: the ratio of the segment assets to the firm
assets.

\emph{Industry q}: the median \(q\) of single-segment firms within the
same industry as the segment.

\emph{Customer size}: the log value of the total assets of the linked
customer company.

\emph{High Customer Size}: equal to one if the customer size is greater
than the median customer size and zero otherwise (the cross-sectional
split of the sample is based on characteristics during the pre-period).

\emph{High Supp Seg HHI}: equals one for supplier segments operating in
three-digit SIC industries with above-median concentration and zero
otherwise (the cross-sectional split of the sample is based on
characteristics during the pre-period).

\newpage
% figures
\section{Figures and Tables}\label{figures-and-tables}

\begin{figure}[H]
\centering
\begin{subfigure}{\textwidth} \centering
\includegraphics[width=0.8\textwidth,height=\textheight]{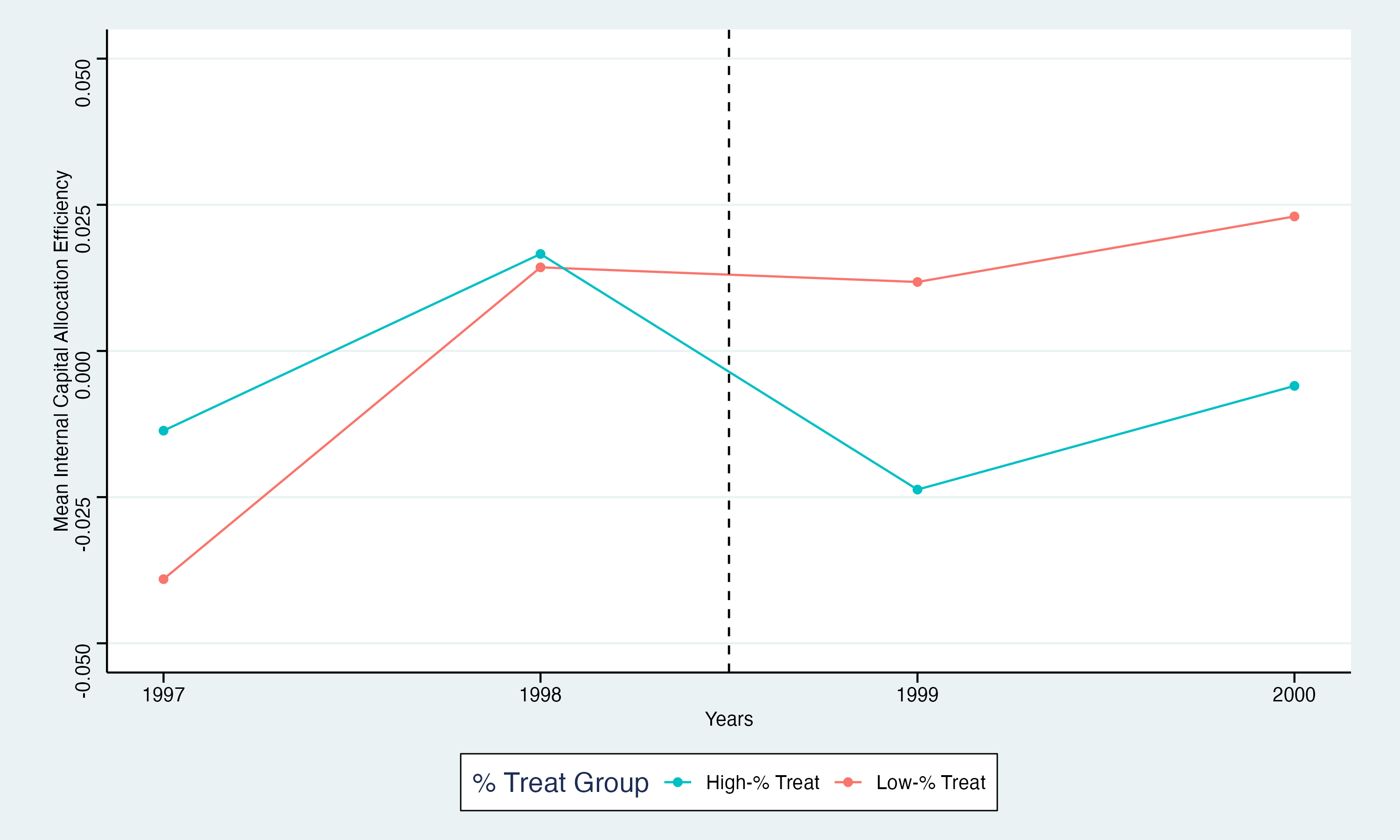}
\caption{Mean Internal Capital Allocation Efficiency for High- and Low-$\% Treat$ Groups}\label{fig:parallel_s14}
\end{subfigure}

\bigskip

\begin{subfigure}{\textwidth} \centering
\includegraphics[width=0.8\textwidth,height=\textheight]{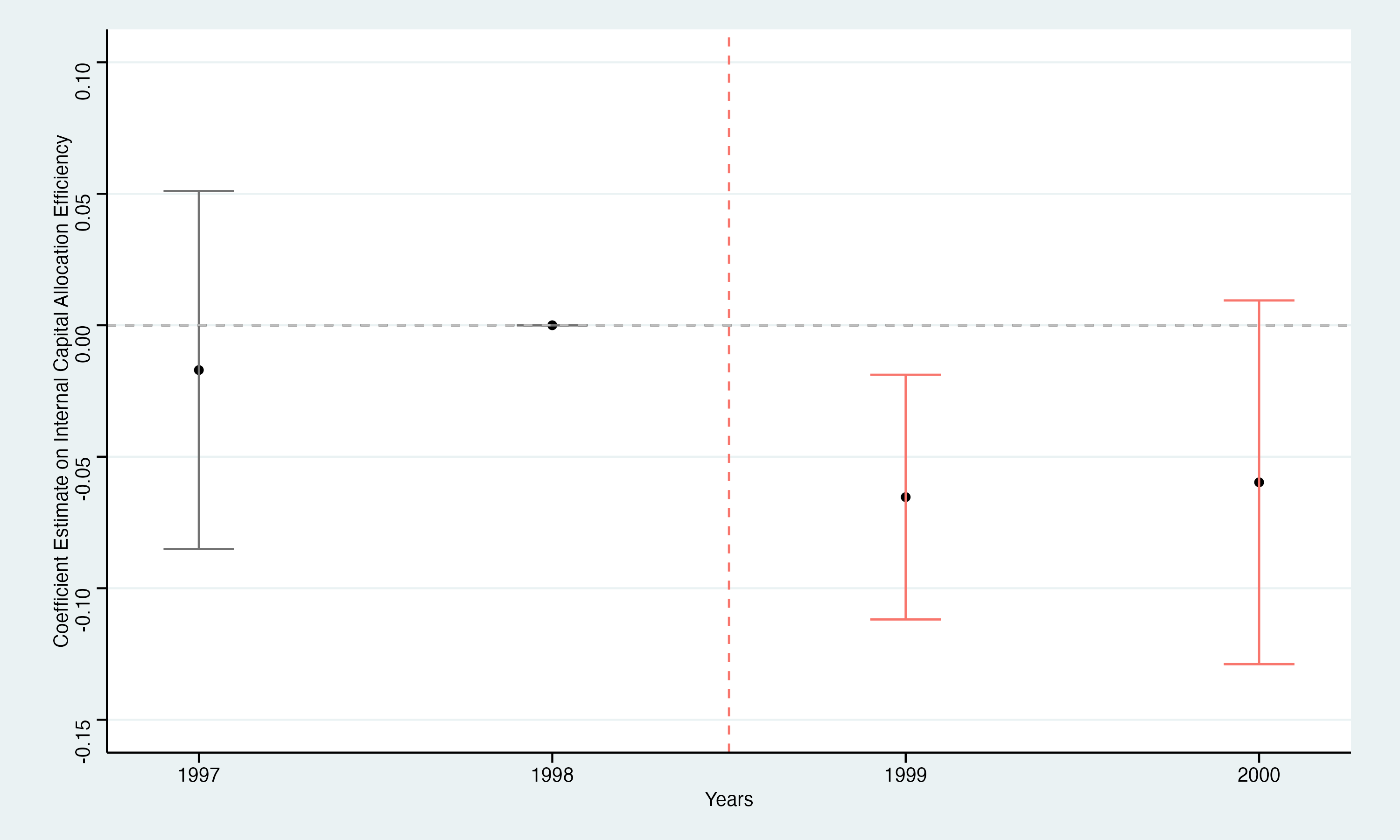}
\caption{Coefficients on Internal Capital Allocation Efficiency}\label{fig:coef_s14_93}
\end{subfigure}

\caption{Parallel Trends in Internal Capital Allocation Efficiency}
\end{figure}

\cref{fig:parallel_s14} plots the average internal capital allocation
efficiency for firms in the high- and low-\(\% Treat\) groups across the
years around the adoption of SFAS 131. 1997 and 1998 on the \(x\)-axis
denote the pre-period, and 1999 and 2000 denote the post-period. The
high-\(\% Treat\) group includes firms with \(\% Treat\) at or above the
sample median. Because the median of \(\% Treat\) is 1, this group
consists of suppliers whose major customers all expanded their segment
disclosures. The low-\(\% Treat\) group includes firms with \(\% Treat\)
below 1.

\cref{fig:coef_s14_93} plots the \(\beta_\tau\) coefficients from
\cref{eq:firm_capital_allocation_event_study}, reproduced here for
reference: \[
\begin{aligned}
Capital\text{ }Allocation\text{ }Efficiency_{it}
={}& \sum_{\tau \in \{1997,1999,2000\}} \beta_\tau
  \left(\% Treat_{it} \times Year_\tau\right) \\
&+ \gamma Controls_{it} + \mathit{Firm\ FE}_i + \mathit{Year\ FE}_t + \varepsilon_{it}.
\end{aligned}
\] where \(Year_\tau\) is an indicator for year \(\tau\). The
\(\mathit{Firm\ FE}_i\) and \(\mathit{Year\ FE}_t\) terms denote firm
and year fixed effects, respectively. \(Year_{1998}\), the year
immediately before the adoption, is omitted for comparison. The figure
plots the 90\% confidence intervals. Standard errors are clustered at
the firm level.

\newpage

\begin{figure}[H]
\centering
\begin{subfigure}{\textwidth} \centering
\includegraphics[width=0.8\textwidth,height=\textheight]{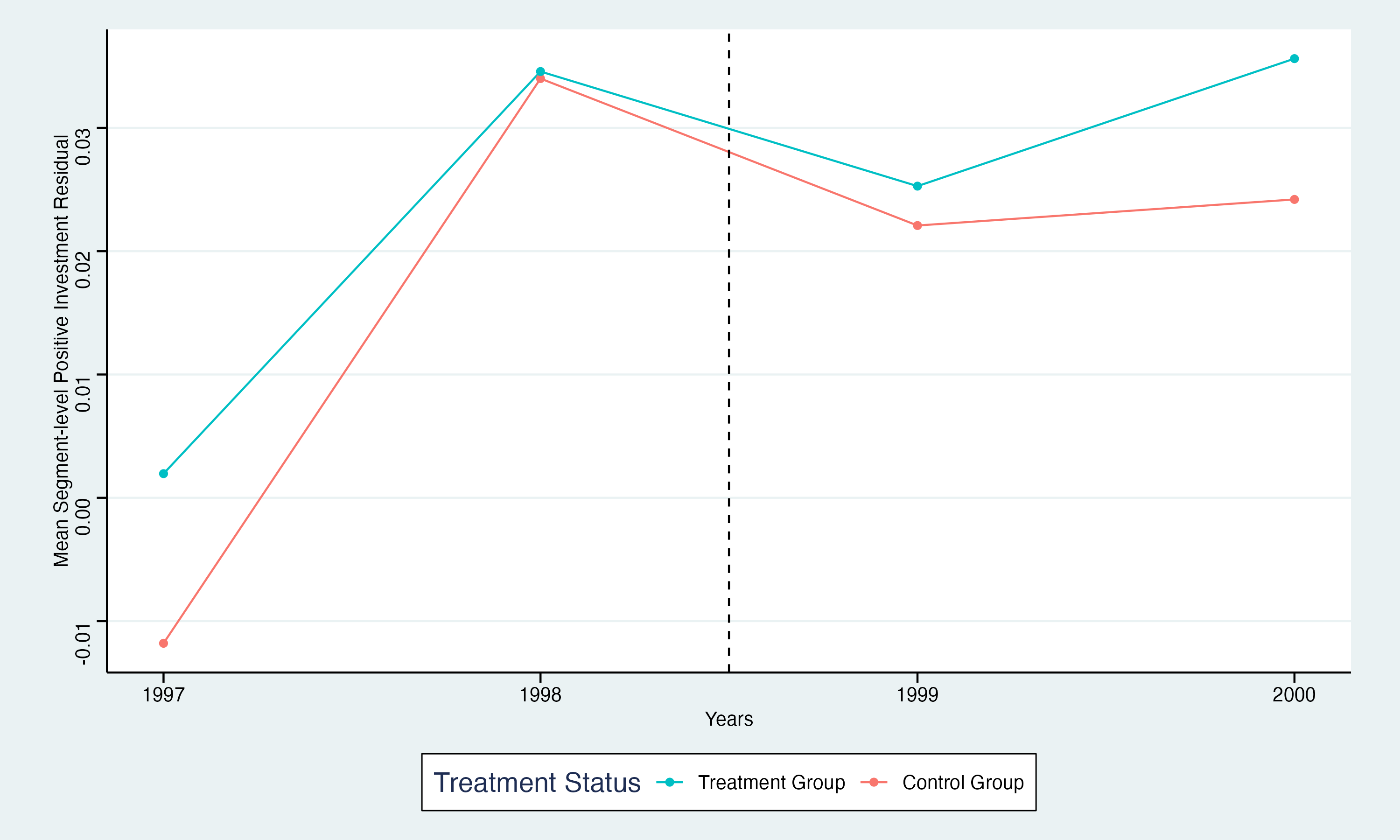}
\caption{Mean Segment-level Positive Investment Residual}\label{fig:parallel_s16}
\end{subfigure}

\bigskip

\begin{subfigure}{\textwidth} \centering
\includegraphics[width=0.8\textwidth,height=\textheight]{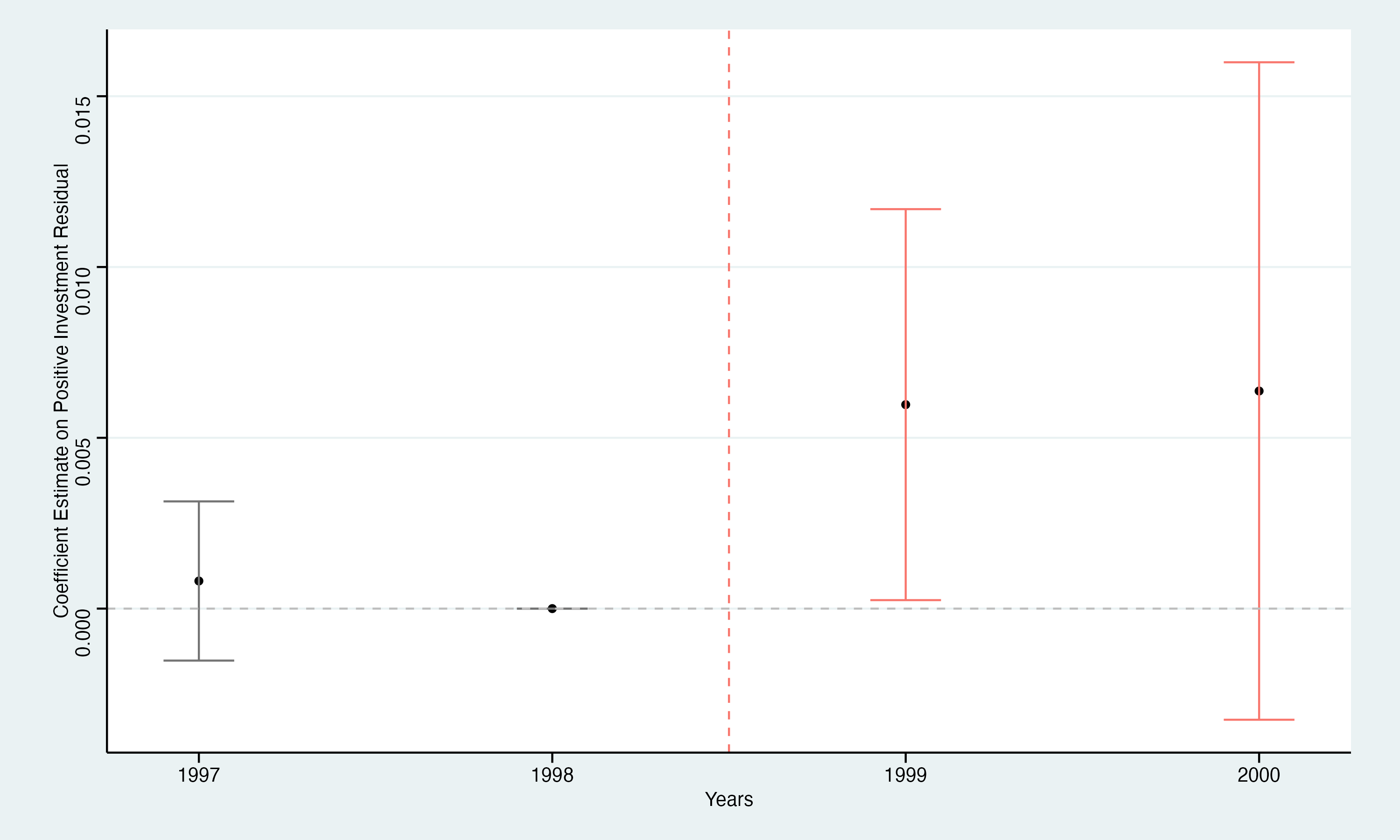}
\caption{Coefficients on Positive Investment Residual}\label{fig:coef_s19_03}
\end{subfigure}

\caption{Parallel Trends in Segment-level Positive Investment Residual}
\end{figure}

\cref{fig:parallel_s16} plots the average value of the \emph{Positive
Investment Residual} variable for both the treated and control groups
across the years around the adoption of SFAS 131. 1997 and 1998 on the
\(x\)-axis denote the pre-period and 1999 and 2000 indicate the years
following it, respectively. The treatment group consists of supplier
segments whose linked customer expanded its segment disclosures. The
control group consists of supplier segments whose linked customer did
not expand its segment disclosures.

\cref{fig:coef_s19_03} plots the \(\beta_\tau\) coefficients from
\cref{eq:segment_investment_event_study}, reproduced here for reference:
\[
\begin{aligned}
Positive\text{ }Investment\text{ }Residual_{ijk,t+1}
={}& \sum_{\tau \in \{1997,1999,2000\}} \beta_\tau
  \left(Treat_{ijk} \times Year_\tau\right) \\
&+ \gamma Controls_{ijkt} + \mathit{Supplier\!-\!Segment\ FE}_{ij} \\
&+ \mathit{Industry\!-\!Year\ FE}_{Industry(j)t} + \varepsilon_{ijkt}.
\end{aligned}
\] where \(Year_\tau\) is an indicator for year \(\tau\). The
\(\mathit{Supplier\!-\!Segment\ FE}_{ij}\) and
\(\mathit{Industry\!-\!Year\ FE}_{Industry(j)t}\) terms denote
supplier-segment and industry-year fixed effects, respectively.
\(Year_{1998}\), the year immediately before the adoption, is omitted
for comparison. The figure plots the 90\% confidence intervals. Standard
errors are clustered at the supplier-segment level.

\newpage

\begin{figure}[H]
\centering
\includegraphics[width=0.8\textwidth,height=\textheight]{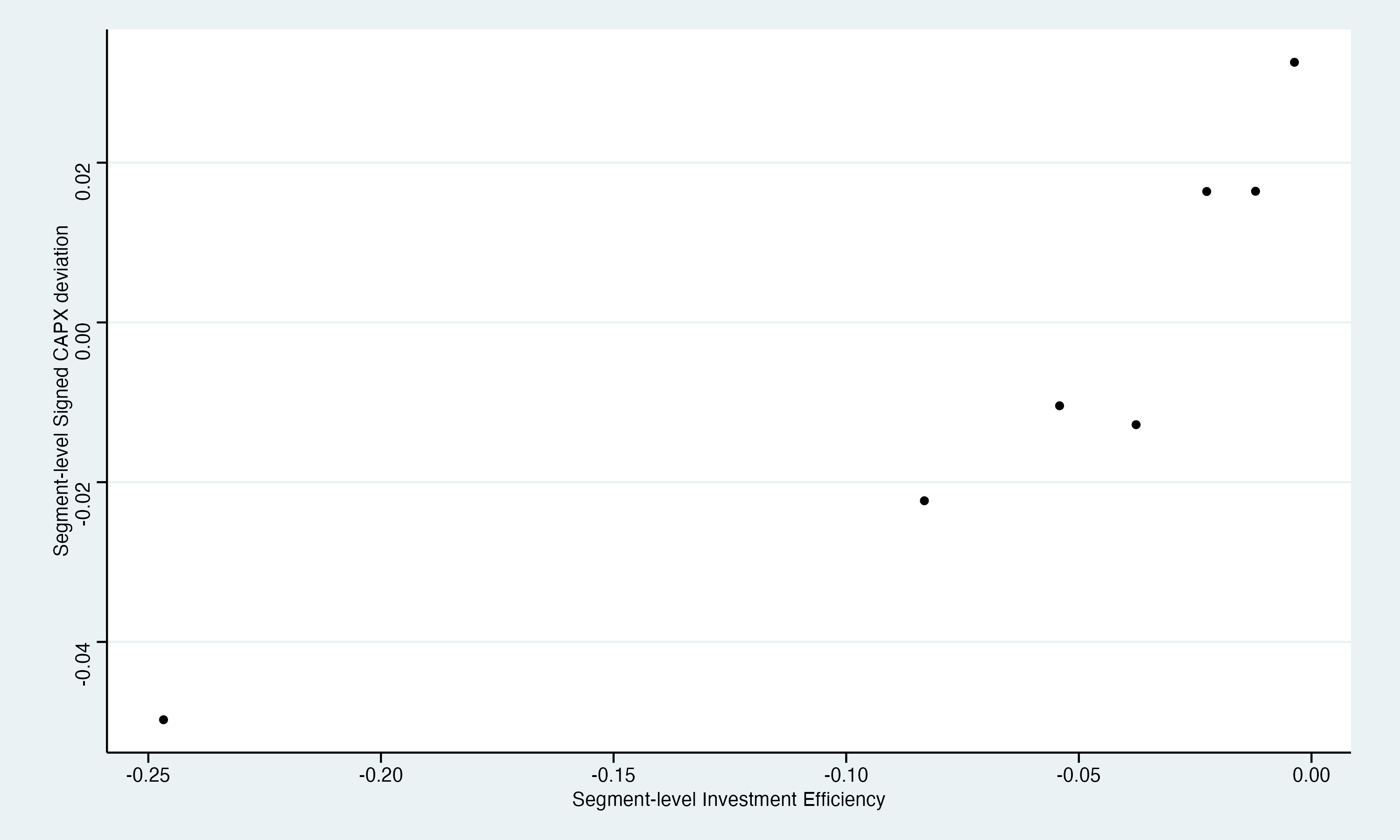}
\caption{Correspondence between Internal Capital Allocation Efficiency Measure and Segment Investment Efficiency Measure}\label{fig:binscatter_s18_08}
\end{figure}

This binned scatter plot depicts the conditional mean function of the
segment-level signed CAPX deviation measure, a component of the
firm-level internal capital allocation efficiency (outlined in
\cref{sec:cho_int_eff}), in relation to the segment-level investment
efficiency measure (defined in \cref{sec:biddle_seg_eff}), controlling
for the segment-level characteristics. The segment-level signed CAPX
deviation measure is increasing in the segment-level investment
efficiency measure. This figure illustrates that the two measures of
investment efficiency used in this study are positively associated with
each other.

\clearpage

\begin{table}[!htbp]

\caption{\label{tbl:s19_01_sample_selection}Sample Selection}
\centering
\begin{tabularx}{\textwidth}{>{\raggedright\arraybackslash}Xrrrr}
\toprule
Sample step & Supplier-years & Attrition & Suppliers & Customers\\
\midrule
Compustat segment data between fiscal years 1996 and 2000 & 47,010 &  & 12,434 & \\
Supplier-years with linked major customers & 9,062 & (37,948) & 3,499 & 2,122\\
Supplier-years excluding real segment-structure changes & 5,480 & (3,582) & 1,803 & 1,408\\
Supplier-years after financial and regulated-industry exclusions & 5,408 & (72) & 1,777 & 1,389\\
Supplier-years with complete segment assets, capital expenditures, and sales & 1,307 & (4,101) & 444 & 543\\
Supplier-years with customers in the SFAS 131 classification sample & 1,141 & (166) & 386 & 476\\
Supplier-years with treatment variables & 902 & (239) & 318 & 390\\
Supplier-years with complete controls & 897 & (5) & 316 & 386\\
\bottomrule
\end{tabularx}

   \caption*{\footnotesize
This table describes the sample selection criteria for the 1996--2000 firm-level sample. Attrition is the decrease in supplier-years from the immediately preceding row. Customers reports the number of unique linked major customers.
   }
\end{table}

\clearpage

% descriptive stat tables
\newgeometry{margin=0.1in}
\begin{landscape}
\begin{table}[!htbp] \centering 
  \caption{Firm-level Descriptive Statistics}
  \label{tbl:desc_firm}
  \footnotesize
  \renewcommand*{\arraystretch}{0.88}
  \caption*{\textbf{Panel A: High-$\% Treat$ Group}}
  \label{tbl:desc_firm_treat}
  \resizebox{0.93\linewidth}{!}{%
  \begin{tabular}{@{\extracolsep{5pt}}lcccccc|cccccc}
    \toprule
   & \multicolumn{6}{c}{Pre-SFAS 131} & \multicolumn{6}{c}{Post-SFAS 131} \\
    Statistic & \multicolumn{1}{c}{N} & \multicolumn{1}{c}{Mean} & \multicolumn{1}{c}{St. Dev.} & \multicolumn{1}{c}{Pctl(25)} & \multicolumn{1}{c}{Median} & \multicolumn{1}{c}{Pctl(75)} & \multicolumn{1}{c}{N} & \multicolumn{1}{c}{Mean} & \multicolumn{1}{c}{St. Dev.} & \multicolumn{1}{c}{Pctl(25)} & \multicolumn{1}{c}{Median} & \multicolumn{1}{c}{Pctl(75)} \\ 
    \hline \\[-1.8ex]
Capital Allocation Efficiency & 221 & 0.001 & 0.195 & $-$0.073 & $-$0.004 & 0.099 & 231 & $-$0.014 & 0.183 & $-$0.099 & $-$0.013 & 0.073 \\ 
\% Treat & 221 & 1.000 & 0.000 & 1 & 1 & 1 & 231 & 1.000 & 0.000 & 1 & 1 & 1 \\ 
N Rivals & 179 & 33.626 & 41.577 & 7 & 18 & 45 & 193 & 30.782 & 41.342 & 5 & 12 & 32 \\ 
Textual Similarities & 207 & 2.206 & 2.359 & 1.091 & 1.357 & 1.988 & 219 & 2.278 & 2.626 & 1.063 & 1.241 & 1.887 \\ 
log(Total Assets) & 221 & 5.796 & 1.927 & 4.469 & 5.795 & 7.149 & 231 & 5.855 & 1.895 & 4.452 & 5.981 & 7.168 \\ 
Market-to-Book & 221 & 2.866 & 3.451 & 1.407 & 2.169 & 3.206 & 231 & 2.552 & 3.536 & 0.957 & 1.506 & 2.828 \\ 
Cashflow & 221 & 0.085 & 0.123 & 0.040 & 0.097 & 0.153 & 231 & 0.084 & 0.138 & 0.032 & 0.087 & 0.157 \\ 
CapEx & 221 & 0.257 & 0.143 & 0.165 & 0.222 & 0.321 & 231 & 0.227 & 0.141 & 0.128 & 0.188 & 0.293 \\ 
NonCapEx & 221 & 0.891 & 0.312 & 1 & 1 & 1 & 231 & 0.892 & 0.311 & 1 & 1 & 1 \\ 
Tangibility & 221 & 0.313 & 0.186 & 0.173 & 0.280 & 0.407 & 231 & 0.329 & 0.188 & 0.194 & 0.288 & 0.430 \\ 
Cash & 221 & 0.098 & 0.123 & 0.015 & 0.044 & 0.134 & 231 & 0.088 & 0.123 & 0.011 & 0.035 & 0.118 \\ 
Leverage & 221 & 0.536 & 0.219 & 0.393 & 0.553 & 0.689 & 231 & 0.536 & 0.229 & 0.369 & 0.549 & 0.681 \\ 
Dividend & 221 & 0.516 & 0.501 & 0 & 1 & 1 & 231 & 0.468 & 0.500 & 0 & 0 & 1 \\ 
External Financing & 221 & 0.674 & 4.592 & $-$0.516 & 0.174 & 1.181 & 231 & 0.933 & 4.826 & $-$1.090 & 0.012 & 1.319 \\ 
N Segment & 221 & 2.792 & 1.001 & 2 & 3 & 3 & 231 & 2.688 & 0.936 & 2 & 2 & 3 \\ 
Speed of Profit Adjustment & 221 & 0.315 & 0.224 & 0.110 & 0.282 & 0.436 & 231 & 0.275 & 0.249 & 0.082 & 0.201 & 0.415 \\ 
Concentration Ratio & 221 & 0.055 & 0.030 & 0.035 & 0.045 & 0.073 & 231 & 0.053 & 0.026 & 0.036 & 0.045 & 0.068 \\ 
Seg Earnings Persistence & 221 & 0.207 & 0.251 & 0.043 & 0.124 & 0.274 & 231 & 0.201 & 0.238 & 0.042 & 0.126 & 0.341 \\ 
Seg Industry Diversity & 221 & 0.735 & 0.269 & 0.500 & 0.667 & 1.000 & 231 & 0.752 & 0.261 & 0.500 & 0.667 & 1.000 \\ 
    \hline \\[-1.8ex] 
  \end{tabular}}

  \caption*{\textbf{Panel B: Low-$\% Treat$ Group}}
  \label{tbl:desc_firm_control}
  \footnotesize
  \resizebox{0.93\linewidth}{!}{%
  \begin{tabular}{@{\extracolsep{5pt}}lcccccc|cccccc}
    \toprule
   & \multicolumn{6}{c}{Pre-SFAS 131} & \multicolumn{6}{c}{Post-SFAS 131} \\
    Statistic & \multicolumn{1}{c}{N} & \multicolumn{1}{c}{Mean} & \multicolumn{1}{c}{St. Dev.} & \multicolumn{1}{c}{Pctl(25)} & \multicolumn{1}{c}{Median} & \multicolumn{1}{c}{Pctl(75)} & \multicolumn{1}{c}{N} & \multicolumn{1}{c}{Mean} & \multicolumn{1}{c}{St. Dev.} & \multicolumn{1}{c}{Pctl(25)} & \multicolumn{1}{c}{Median} & \multicolumn{1}{c}{Pctl(75)} \\ 
    \hline \\[-1.8ex]
Capital Allocation Efficiency & 202 & $-$0.012 & 0.268 & $-$0.109 & $-$0.003 & 0.107 & 243 & 0.018 & 0.259 & $-$0.108 & 0.004 & 0.149 \\ 
\% Treat & 202 & 0.219 & 0.259 & 0.000 & 0.000 & 0.500 & 243 & 0.269 & 0.268 & 0.000 & 0.333 & 0.500 \\ 
N Rivals & 166 & 30.470 & 41.705 & 4 & 12 & 34 & 184 & 30.734 & 42.559 & 5 & 13 & 34.5 \\ 
Textual Similarities & 193 & 2.240 & 2.505 & 1.057 & 1.246 & 1.949 & 224 & 2.032 & 2.154 & 1.037 & 1.200 & 1.883 \\ 
log(Total Assets) & 202 & 4.956 & 1.584 & 3.942 & 4.851 & 5.821 & 243 & 5.146 & 1.713 & 3.939 & 5.067 & 6.295 \\ 
Market-to-Book & 202 & 3.434 & 4.488 & 1.443 & 2.242 & 3.355 & 243 & 3.343 & 5.523 & 0.837 & 1.431 & 3.124 \\ 
Cashflow & 202 & 0.082 & 0.148 & 0.008 & 0.079 & 0.148 & 243 & 0.052 & 0.133 & $-$0.002 & 0.052 & 0.115 \\ 
CapEx & 202 & 0.277 & 0.170 & 0.154 & 0.244 & 0.366 & 243 & 0.235 & 0.160 & 0.124 & 0.191 & 0.292 \\ 
NonCapEx & 202 & 0.752 & 0.433 & 1 & 1 & 1 & 243 & 0.798 & 0.402 & 1 & 1 & 1 \\ 
Tangibility & 202 & 0.310 & 0.197 & 0.150 & 0.284 & 0.395 & 243 & 0.304 & 0.191 & 0.148 & 0.271 & 0.427 \\ 
Cash & 202 & 0.124 & 0.156 & 0.020 & 0.056 & 0.150 & 243 & 0.102 & 0.150 & 0.014 & 0.038 & 0.109 \\ 
Leverage & 202 & 0.490 & 0.230 & 0.309 & 0.450 & 0.646 & 243 & 0.551 & 0.267 & 0.347 & 0.521 & 0.734 \\ 
Dividend & 202 & 0.391 & 0.489 & 0 & 0 & 1 & 243 & 0.342 & 0.475 & 0 & 0 & 1 \\ 
External Financing & 202 & 0.840 & 4.659 & $-$0.601 & 0.146 & 1.909 & 243 & 0.379 & 5.701 & $-$0.971 & 0.052 & 1.445 \\ 
N Segment & 202 & 2.856 & 1.095 & 2 & 3 & 3 & 243 & 2.831 & 0.992 & 2 & 3 & 3 \\ 
Speed of Profit Adjustment & 202 & 0.339 & 0.243 & 0.141 & 0.325 & 0.479 & 243 & 0.296 & 0.248 & 0.086 & 0.251 & 0.452 \\ 
Concentration Ratio & 202 & 0.059 & 0.033 & 0.037 & 0.046 & 0.077 & 243 & 0.060 & 0.034 & 0.037 & 0.050 & 0.074 \\ 
Seg Earnings Persistence & 202 & 0.170 & 0.208 & 0.038 & 0.090 & 0.240 & 243 & 0.196 & 0.290 & 0.034 & 0.095 & 0.281 \\ 
Seg Industry Diversity & 202 & 0.745 & 0.278 & 0.500 & 0.833 & 1.000 & 243 & 0.737 & 0.276 & 0.500 & 0.750 & 1.000 \\ 
    \hline \\[-1.8ex] 
  \end{tabular}}
   \caption*{\footnotesize
This table presents descriptive statistics for firm-level variables for the High-$\% Treat$ and Low-$\% Treat$ groups in the pre- and post-SFAS 131 periods. The High-$\% Treat$ group includes firms with $\% Treat$ at or above the sample median, and the Low-$\% Treat$ group includes firms with $\% Treat$ below the sample median. All variable definitions are in \cref{sec:var_def}. All continuous variables are winsorized at the 1\% and 99\% levels.
   }
\end{table}
\end{landscape}
\restoregeometry

\clearpage
\begin{landscape}
\begin{table}[!htbp] \centering 
  \caption{Segment-level Descriptive Statistics}
  \label{tbl:desc_seg}
  \footnotesize
  \renewcommand*{\arraystretch}{0.95}
  \caption*{\textbf{Panel A: Treatment Group}}
  \label{tbl:desc_seg_treat}
  \resizebox{0.95\linewidth}{!}{%
  \begin{tabular}{@{\extracolsep{5pt}}lccccc|ccccc}
    \toprule
    & \multicolumn{5}{c}{Treatment (Pre, N = 1,288)} & \multicolumn{5}{c}{Treatment (Post, N = 1,239)} \\
    Statistic & \multicolumn{1}{c}{Mean} & \multicolumn{1}{c}{St. Dev.} & \multicolumn{1}{c}{Pctl(25)} & \multicolumn{1}{c}{Median} & \multicolumn{1}{c}{Pctl(75)} & \multicolumn{1}{c}{Mean} & \multicolumn{1}{c}{St. Dev.} & \multicolumn{1}{c}{Pctl(25)} & \multicolumn{1}{c}{Median} & \multicolumn{1}{c}{Pctl(75)} \\ 
    \hline \\[-1.8ex]
InvEff (Abs) & $-$0.078 & 0.117 & $-$0.088 & $-$0.038 & $-$0.016 & $-$0.062 & 0.096 & $-$0.068 & $-$0.039 & $-$0.015 \\ 
Positive Inv Resid & 0.031 & 0.079 & 0.000 & 0.000 & 0.020 & 0.029 & 0.083 & 0.000 & 0.000 & 0.017 \\ 
Negative Inv Resid & 0.040 & 0.065 & 0.000 & 0.014 & 0.055 & 0.031 & 0.045 & 0.000 & 0.014 & 0.046 \\ 
Seg Size & 4.317 & 1.820 & 2.935 & 4.324 & 5.608 & 4.296 & 2.086 & 2.499 & 4.602 & 5.924 \\ 
Seg Relative Size & 0.392 & 0.285 & 0.143 & 0.306 & 0.643 & 0.398 & 0.280 & 0.172 & 0.320 & 0.592 \\ 
Industry $q$ & 1.910 & 0.682 & 1.478 & 1.655 & 2.144 & 1.904 & 1.012 & 1.179 & 1.415 & 2.557 \\ 
Customer Sales Growth & 0.247 & 0.923 & $-$0.005 & 0.049 & 0.137 & 0.167 & 0.701 & $-$0.032 & 0.066 & 0.160 \\ 
Seg Sales Growth & 0.609 & 1.933 & 0.031 & 0.156 & 0.389 & 0.718 & 2.431 & $-$0.017 & 0.086 & 0.424 \\ 
Customer Size & 11.189 & 1.552 & 10.296 & 12.341 & 12.378 & 10.648 & 1.776 & 9.774 & 10.506 & 12.458 \\ 
\hline \\[-1.8ex] 
\end{tabular}} 
  \caption*{\textbf{Panel B: Control Group}}
  \label{tbl:desc_seg_control}
  \footnotesize
  \resizebox{0.95\linewidth}{!}{%
  \begin{tabular}{@{\extracolsep{5pt}}lccccc|ccccc}
  \toprule
    & \multicolumn{5}{c}{Control (Pre, N = 729)} & \multicolumn{5}{c}{Control (Post, N = 1,189)} \\
    Statistic & \multicolumn{1}{c}{Mean} & \multicolumn{1}{c}{St. Dev.} & \multicolumn{1}{c}{Pctl(25)} & \multicolumn{1}{c}{Median} & \multicolumn{1}{c}{Pctl(75)} & \multicolumn{1}{c}{Mean} & \multicolumn{1}{c}{St. Dev.} & \multicolumn{1}{c}{Pctl(25)} & \multicolumn{1}{c}{Median} & \multicolumn{1}{c}{Pctl(75)} \\ 
    \hline \\[-1.8ex]
InvEff (Abs) & $-$0.070 & 0.110 & $-$0.077 & $-$0.037 & $-$0.013 & $-$0.056 & 0.074 & $-$0.065 & $-$0.036 & $-$0.015 \\ 
Positive Inv Resid & 0.036 & 0.089 & 0.000 & 0.000 & 0.018 & 0.024 & 0.060 & 0.000 & 0.000 & 0.022 \\ 
Negative Inv Resid & 0.030 & 0.049 & 0.000 & 0.011 & 0.047 & 0.031 & 0.050 & 0.000 & 0.011 & 0.042 \\ 
Seg Size & 3.918 & 1.640 & 2.795 & 3.959 & 4.985 & 3.904 & 1.932 & 2.285 & 4.155 & 5.221 \\ 
Seg Relative Size & 0.380 & 0.287 & 0.137 & 0.278 & 0.602 & 0.379 & 0.272 & 0.147 & 0.302 & 0.577 \\ 
Industry $q$ & 1.857 & 0.724 & 1.255 & 1.626 & 2.207 & 1.762 & 0.923 & 1.179 & 1.338 & 2.125 \\ 
Customer Sales Growth & 0.164 & 0.373 & 0.023 & 0.082 & 0.163 & 0.166 & 0.421 & 0.004 & 0.088 & 0.203 \\ 
Seg Sales Growth & 0.399 & 1.560 & $-$0.011 & 0.100 & 0.256 & 0.260 & 1.200 & $-$0.066 & 0.075 & 0.243 \\ 
Customer Size & 9.915 & 1.191 & 9.537 & 10.214 & 10.723 & 10.147 & 1.470 & 9.659 & 10.321 & 10.820 \\ 
\hline \\[-1.8ex] 

\end{tabular}} 
   \caption*{\footnotesize
This table presents descriptive statistics for segment-level variables related to Treated and Control segments across both pre-SFAS 131 and post-SFAS 131 periods. All variable definitions are in \cref{sec:var_def}. All continuous variables are winsorized at the 1\% and 99\% levels.
   }
\end{table} 
\end{landscape}

\clearpage

% n rivals
\begin{table}[htbp]
   \caption{\label{tbl:n_rivals_competition} (Firm-level Analysis) Customer Disclosures and Supplier-side Competition}
   \centering
   \scriptsize
   \begin{adjustbox}{width = \textwidth, center}
      \renewcommand*{\arraystretch}{1.02}
      \begin{tabular}{lcccccccc}
         \tabularnewline \midrule \midrule
         Dependent Variables: & \multicolumn{4}{c}{N Rivals (Hoberg-Phillips)} & \multicolumn{4}{c}{Textual Similarities (Hoberg-Phillips)}\\
         Model:                     & (1)           & (2)           & (3)          & (4)           & (5)            & (6)            & (7)           & (8)\\  
         \midrule
         \emph{Variables}\\
   \rowcolor{Gainsboro!60} % \rowcolor{lightgray}
         \% Treat x Post            & 7.357$^{***}$ & 7.749$^{***}$ & 7.702$^{**}$ & 9.085$^{***}$ & 0.2804$^{***}$ & 0.2873$^{***}$ & 0.2331$^{**}$ & 0.2826$^{**}$\\   
   \rowcolor{Gainsboro!60} % \rowcolor{lightgray}
                                    & (3.372)       & (3.240)       & (2.591)      & (2.768)       & (3.081)        & (2.727)        & (2.139)       & (2.380)\\   
         log(Total Assets)          &               & 10.41$^{***}$ &              & 8.654$^{*}$   &                & 0.3134$^{**}$  &               & 0.4036$^{**}$\\   
                                    &               & (2.780)       &              & (1.685)       &                & (2.156)        &               & (2.199)\\   
         Market-to-Book             &               & -0.2159       &              & -0.2689       &                & -0.0090        &               & -0.0088\\   
                                    &               & (-1.024)      &              & (-0.8118)     &                & (-1.026)       &               & (-0.8181)\\   
         Cashflow                   &               & -19.56        &              & -31.35$^{**}$ &                & -0.9084$^{*}$  &               & -0.8807$^{*}$\\   
                                    &               & (-1.541)      &              & (-2.424)      &                & (-1.822)       &               & (-1.754)\\   
         CapEx                      &               & 8.325         &              & 2.145         &                & 0.2332         &               & 0.2175\\   
                                    &               & (1.138)       &              & (0.2409)      &                & (0.9582)       &               & (0.5765)\\   
         NonCapEx                   &               & -3.138        &              & -2.146        &                & 0.2061         &               & 0.1637\\   
                                    &               & (-0.5956)     &              & (-0.3518)     &                & (0.6550)       &               & (0.7902)\\   
         Tangibility                &               & -9.116        &              & -18.63        &                & 0.9997         &               & 1.030\\   
                                    &               & (-0.4202)     &              & (-0.7436)     &                & (0.9780)       &               & (1.106)\\   
         Cash                       &               & 18.41         &              & -2.480        &                & 1.852$^{**}$   &               & 0.7992\\   
                                    &               & (1.062)       &              & (-0.1596)     &                & (2.255)        &               & (0.9340)\\   
         Leverage                   &               & 13.76$^{**}$  &              & 2.893         &                & 0.2921         &               & -0.0055\\   
                                    &               & (2.465)       &              & (0.3943)      &                & (1.276)        &               & (-0.0186)\\   
         Dividend                   &               & 1.123         &              & 3.728         &                & 0.0236         &               & 0.0452\\   
                                    &               & (0.5742)      &              & (1.081)       &                & (0.1994)       &               & (0.3101)\\   
         External Financing         &               & -0.1738       &              & -0.0109       &                & -0.0137        &               & -0.0158\\   
                                    &               & (-0.9440)     &              & (-0.0540)     &                & (-1.202)       &               & (-1.173)\\   
         N Segment                  &               & -1.774        &              & -3.379        &                & -0.1446$^{**}$ &               & -0.1188\\   
                                    &               & (-1.078)      &              & (-0.8504)     &                & (-2.060)       &               & (-1.321)\\   
         Speed of Profit Adjustment &               & 8.274         &              & -13.03        &                & 0.3354         &               & -0.3844\\   
                                    &               & (1.438)       &              & (-0.8841)     &                & (1.485)        &               & (-1.104)\\   
         Concentration Ratio        &               & -33.32        &              & -28.45        &                & 1.792          &               & 3.152\\   
                                    &               & (-0.4351)     &              & (-0.1465)     &                & (0.8693)       &               & (0.5916)\\   
         Seg Earnings Persistence   &               & 1.813         &              & 2.667         &                & 0.1892         &               & 0.2649\\   
                                    &               & (0.6399)      &              & (0.2700)      &                & (1.607)        &               & (1.194)\\   
         Seg Industry Diversity     &               & -10.75        &              & -25.60        &                & -0.2238        &               & -0.7826$^{*}$\\   
                                    &               & (-1.158)      &              & (-1.601)      &                & (-0.8101)      &               & (-1.914)\\   
         \midrule
         \emph{Fixed-effects}\\
         Firm                       & Yes           & Yes           & Yes          & Yes           & Yes            & Yes            & Yes           & Yes\\  
         Year                       & Yes           & Yes           &              &               & Yes            & Yes            &               & \\  
         Industry-Year              &               &               & Yes          & Yes           &                &                & Yes           & Yes\\  
         \midrule
         \emph{Fit statistics}\\
         Observations               & 722           & 722           & 722          & 722           & 843            & 843            & 843           & 843\\  
         R$^2$                      & 0.94148       & 0.94583       & 0.96029      & 0.96426       & 0.96258        & 0.96480        & 0.97382       & 0.97504\\  
         Adjusted R$^2$             & 0.90300       & 0.90700       & 0.90128      & 0.90629       & 0.94067        & 0.94255        & 0.94152       & 0.94193\\  
         \midrule \midrule
         \multicolumn{9}{l}{\emph{Clustered (Firm) co-variance matrix, t-stats in parentheses}}\\
         \multicolumn{9}{l}{\emph{Signif. Codes: ***: 0.01, **: 0.05, *: 0.1}}\\
      \end{tabular}
   \end{adjustbox}

   \caption*{\footnotesize
This table presents difference-in-differences (DiD) regression results for the effect of customer disclosures on supplier-side competition. I measure competition using two measures based on the Text-based Network Industry Classifications (TNIC-3; see \textcite{hoberg-phillips-2010-product} and \textcite{hoberg-phillips-2016-textbased}). The \emph{N Rivals} measure is the number of rivals in the same TNIC-3 industry, and the \emph{Textual Similarities} measure is the textual similarity of product descriptions with those rivals. All variable definitions are in \cref{sec:var_def}. The unit of observation is the firm-year. Columns (1), (2), (5), and (6) include firm and year fixed effects. Columns (3), (4), (7), and (8) include firm and industry-year fixed effects, where industries are defined at the two-digit SIC level. $t$-statistics are reported in parentheses. Standard errors are clustered at the firm level. All continuous variables are winsorized at the 1\% and 99\% levels.
   }
\end{table}

\clearpage

% main results
\begin{table}[htbp]
   \caption{\label{tbl:main_results_firm} (Firm-level Analysis) Customer Disclosures and Suppliers' Internal Capital Allocation}
   \centering
   \scriptsize
   \begin{adjustbox}{width = 0.82\textwidth, center}
      \renewcommand*{\arraystretch}{0.68}
      \begin{tabular}{lcccc}
         \tabularnewline \midrule \midrule
         Dependent Variable: & \multicolumn{4}{c}{Capital Allocation Efficiency}\\
         Model:                     & (1)            & (2)            & (3)           & (4)\\  
         \midrule
         \emph{Variables}\\
   \rowcolor{Gainsboro!60} % \rowcolor{lightgray}
         \% Treat x Post            & -0.0586$^{**}$ & -0.0614$^{**}$ & -0.0578$^{*}$ & -0.0593$^{*}$\\   
   \rowcolor{Gainsboro!60} % \rowcolor{lightgray}
                                    & (-2.273)       & (-2.190)       & (-1.674)      & (-1.698)\\   
         log(Total Assets)          &                & 0.0319         &               & 0.0740\\   
                                    &                & (0.8734)       &               & (1.441)\\   
         Market-to-Book             &                & -0.0011        &               & 0.0011\\   
                                    &                & (-0.1640)      &               & (0.1408)\\   
         Cashflow                   &                & -0.0224        &               & -0.0356\\   
                                    &                & (-0.2709)      &               & (-0.3224)\\   
         CapEx                      &                & 0.0445         &               & 0.0459\\   
                                    &                & (0.4534)       &               & (0.3816)\\   
         NonCapEx                   &                & 0.0103         &               & -0.0359\\   
                                    &                & (0.2194)       &               & (-0.5265)\\   
         Tangibility                &                & -0.1276        &               & -0.1414\\   
                                    &                & (-0.7571)      &               & (-0.7040)\\   
         Cash                       &                & 0.0384         &               & 0.0347\\   
                                    &                & (0.2677)       &               & (0.2226)\\   
         Leverage                   &                & -0.0698        &               & -0.0984\\   
                                    &                & (-0.9580)      &               & (-1.003)\\   
         Dividend                   &                & 0.0043         &               & -0.0252\\   
                                    &                & (0.1109)       &               & (-0.7175)\\   
         External Financing         &                & 0.0005         &               & -0.0007\\   
                                    &                & (0.1486)       &               & (-0.1519)\\   
         N Segment                  &                & 0.0176         &               & 0.0315\\   
                                    &                & (0.5045)       &               & (0.6637)\\   
         Speed of Profit Adjustment &                & 0.0394         &               & -0.1121\\   
                                    &                & (0.4803)       &               & (-0.6706)\\   
         Concentration Ratio        &                & -1.554$^{*}$   &               & -0.8179\\   
                                    &                & (-1.728)       &               & (-0.4965)\\   
         Seg Earnings Persistence   &                & -0.1168        &               & 0.0241\\   
                                    &                & (-1.016)       &               & (0.1895)\\   
         Seg Industry Diversity     &                & -0.0360        &               & -0.0507\\   
                                    &                & (-0.2941)      &               & (-0.3591)\\   
         \midrule
         \emph{Fixed-effects}\\
         Firm                       & Yes            & Yes            & Yes           & Yes\\  
         Year                       & Yes            & Yes            &               & \\  
         Industry-Year              &                &                & Yes           & Yes\\  
         \midrule
         \emph{Fit statistics}\\
         Observations               & 897            & 897            & 897           & 897\\  
         R$^2$                      & 0.54146        & 0.55137        & 0.70464       & 0.71170\\  
         Adjusted R$^2$             & 0.28547        & 0.28219        & 0.35923       & 0.35096\\  
         \midrule \midrule
         \multicolumn{5}{l}{\emph{Clustered (Firm) co-variance matrix, t-stats in parentheses}}\\
         \multicolumn{5}{l}{\emph{Signif. Codes: ***: 0.01, **: 0.05, *: 0.1}}\\
      \end{tabular}
   \end{adjustbox}

   \caption*{\footnotesize
This table presents difference-in-differences (DiD) regression results for the effect of customer disclosures on suppliers' internal capital allocations. All variable definitions are in \cref{sec:var_def}. The unit of observation is the firm-year. Columns (1) and (2) include firm and year fixed effects. Columns (3) and (4) include firm and industry-year fixed effects, where industries are defined at the two-digit SIC level. $t$-statistics are reported in parentheses. Standard errors are clustered at the firm level. All continuous variables are winsorized at the 1\% and 99\% levels.
   }
\end{table}

\clearpage
\begin{table}[htbp]
   \caption{\label{tbl:conseq_01_firm_customer} (Firm-level Analysis) Consequences of Internal Capital Allocation: Market Share and Customer Base}
   \centering
   \footnotesize
   \begin{adjustbox}{width = 0.95\textwidth, center}
      \renewcommand*{\arraystretch}{1.05}
      \begin{tabular}{lcccc}
         \tabularnewline \midrule \midrule
         Dependent Variables:                          & Market Share (t+1) & Market Share (t+2) & N Customers (t+1) & N Customers (t+2)\\  
         Model:                                        & (1)                & (2)                & (3)               & (4)\\  
         \midrule
         \emph{Variables}\\
   \rowcolor{Gainsboro!60} % \rowcolor{lightgray}
         Post $\times$ Capital Reallocation Magnitude  & 0.0285             & 0.1716$^{**}$      & 0.0066            & 0.8696$^{**}$\\   
   \rowcolor{Gainsboro!60} % \rowcolor{lightgray}
                                                       & (0.3641)           & (2.115)            & (0.0176)          & (2.273)\\   
         Capital Reallocation Magnitude                & -0.0503            & -0.1465$^{*}$      & -0.0368           & -0.3393\\   
                                                       & (-0.6449)          & (-1.900)           & (-0.1766)         & (-1.187)\\   
         Post                                          & -0.0064            & 0.0042             & 0.1942            & 0.0822\\   
                                                       & (-0.2278)          & (0.1702)           & (0.9556)          & (0.3276)\\   
         log(Total Assets)                             & -0.0190            & -0.0101            & 0.1744            & 0.1349\\   
                                                       & (-0.6163)          & (-0.1880)          & (0.5638)          & (0.4102)\\   
         Market-to-Book                                & 0.0002             & -0.0023            & 0.0301$^{**}$     & 0.0339$^{**}$\\   
                                                       & (0.0901)           & (-0.9418)          & (2.120)           & (2.255)\\   
         Cashflow                                      & -0.1085            & 0.0101             & 1.104$^{*}$       & 0.0206\\   
                                                       & (-1.236)           & (0.1294)           & (1.714)           & (0.0281)\\   
         CapEx                                         & 0.0995             & 0.0046             & 0.5546            & 0.8907$^{*}$\\   
                                                       & (1.426)            & (0.0785)           & (1.436)           & (1.721)\\   
         NonCapEx                                      & -0.0068            & 0.0085             & 0.5076$^{**}$     & 0.8148$^{***}$\\   
                                                       & (-0.1519)          & (0.1786)           & (2.308)           & (2.994)\\   
         Tangibility                                   & 0.0610             & 0.1405             & 0.8982            & 1.819\\   
                                                       & (0.4358)           & (0.7481)           & (0.5693)          & (1.080)\\   
         Cash                                          & -0.0503            & 0.0383             & 0.7367            & 0.5513\\   
                                                       & (-0.2421)          & (0.2058)           & (0.8658)          & (0.5805)\\   
         Leverage                                      & -0.0799            & -0.0549            & 0.2556            & 0.0909\\   
                                                       & (-1.513)           & (-0.7360)          & (0.7446)          & (0.1584)\\   
         Dividend                                      & -0.0390            & 0.0037             & -0.2669           & -0.1084\\   
                                                       & (-1.282)           & (0.1441)           & (-1.434)          & (-0.5185)\\   
         External Financing                            & -0.0005            & -0.0011            & -0.0023           & 0.0019\\   
                                                       & (-0.2935)          & (-0.5271)          & (-0.1911)         & (0.1810)\\   
         N Segment                                     & -0.0413            & -0.0275            & -0.1136           & 0.0335\\   
                                                       & (-1.174)           & (-0.9850)          & (-0.8540)         & (0.1037)\\   
         \midrule
         \emph{Fixed-effects}\\
         Firm                                          & Yes                & Yes                & Yes               & Yes\\  
         Year                                          & Yes                & Yes                & Yes               & Yes\\  
         \midrule
         \emph{Fit statistics}\\
         Observations                                  & 859                & 804                & 740               & 625\\  
         R$^2$                                         & 0.89102            & 0.90268            & 0.78334           & 0.71387\\  
         Adjusted R$^2$                                & 0.82325            & 0.83821            & 0.63938           & 0.50404\\  
         \midrule \midrule
         \multicolumn{5}{l}{\emph{Clustered (Firm) co-variance matrix, t-stats in parentheses}}\\
         \multicolumn{5}{l}{\emph{Signif. Codes: ***: 0.01, **: 0.05, *: 0.1}}\\
      \end{tabular}
   \end{adjustbox}

   \caption*{\footnotesize
This table presents OLS regression results for the association between firm-level capital reallocation magnitude and market share (Columns (1) and (2)) and the number of customers (Columns (3) and (4)) in subsequent years. The \emph{Capital Reallocation Magnitude} variable equals the \emph{Capital Allocation Efficiency} measure multiplied by -1, so higher values indicate larger deviations from the benchmark. The outcome variable \emph{Market Share} is the ratio of a company's sales to total sales in the same three-digit SIC industry, and \emph{N Customers} is the number of linked customers. All variable definitions are in \cref{sec:var_def}. The unit of observation is the firm-year. I include firm and year fixed effects. $t$-statistics are reported in parentheses. Standard errors are clustered at the firm level. All continuous variables are winsorized at the 1\% and 99\% levels.
   }
\end{table}

\clearpage

\begin{table}[htbp]
   \caption{\label{tbl:main_results_seg} (Segment-level Analysis) Customer Disclosures and Supplier-Segment Investment}
   \centering
   \footnotesize
   \renewcommand*{\arraystretch}{1.08}
   \begin{tabular}{lccc}
      \tabularnewline \midrule \midrule
      Dependent Variables:  & InvEff (Abs)   & Positive Inv Resid & Negative Inv Resid\\  
      Model:                & (1)            & (2)                & (3)\\  
      \midrule
      \emph{Variables}\\
   \rowcolor{Gainsboro!60} % \rowcolor{lightgray}
      Treat x Post          & -0.0089$^{**}$ & 0.0093$^{**}$      & -0.0004\\   
   \rowcolor{Gainsboro!60} % \rowcolor{lightgray}
                            & (-2.017)       & (2.050)            & (-0.2619)\\   
      Seg Sales Growth      & 0.0065$^{**}$  & 0.0004             & -0.0068$^{*}$\\   
                            & (1.972)        & (0.2017)           & (-1.889)\\   
      Customer Sales Growth & 0.0030$^{**}$  & -0.0030$^{***}$    & $-1.5\times 10^{-5}$\\    
                            & (2.548)        & (-2.782)           & (-0.0568)\\   
      Seg Size              & 0.0393         & -0.0578$^{**}$     & 0.0186$^{**}$\\   
                            & (1.338)        & (-2.015)           & (2.328)\\   
      Seg Relative Size     & 0.1152         & -0.1855            & 0.0703\\   
                            & (0.6663)       & (-1.031)           & (1.639)\\   
      Industry q            & -0.0014        & 0.0023             & -0.0009\\   
                            & (-0.0968)      & (0.1472)           & (-0.1842)\\   
      Customer Size         & 0.0044         & -0.0036            & -0.0007\\   
                            & (1.274)        & (-1.080)           & (-0.8524)\\   
      \midrule
      \emph{Fixed-effects}\\
      Firm-segment          & Yes            & Yes                & Yes\\  
      Industry-Year         & Yes            & Yes                & Yes\\  
      \midrule
      \emph{Fit statistics}\\
      Observations          & 4,445          & 4,445              & 4,445\\  
      R$^2$                 & 0.80668        & 0.77707            & 0.89205\\  
      Adjusted R$^2$        & 0.77103        & 0.73595            & 0.87214\\  
      \midrule \midrule
      \multicolumn{4}{l}{\emph{Clustered (Firm-segment) co-variance matrix, t-stats in parentheses}}\\
      \multicolumn{4}{l}{\emph{Signif. Codes: ***: 0.01, **: 0.05, *: 0.1}}\\
   \end{tabular}

   \caption*{\footnotesize
This table presents DiD regression results for the effect of customer disclosures on supplier-segment investments. All variable definitions are in \cref{sec:var_def}. The unit of observation is the supplier segment-customer segment-year. I include firm-segment and industry-year fixed effects following \textcite{chen-etal-2018-incorrect} and \textcite{chen-etal-2022-use}. $t$-statistics are reported in parentheses. Standard errors are clustered at the firm-segment level. All continuous variables are winsorized at the 1\% and 99\% levels.
   }
\end{table}

\clearpage
%\input{tables/s16_03_hidden}

% cross-sections
\begin{table}[htbp]
   \caption{\label{tbl:xsect_seg_01_csize} (Segment-level Analysis) Customer Disclosures and Supplier-Segment Investment: The Role of Customer Importance}
   \centering
   \footnotesize
   \renewcommand*{\arraystretch}{1.05}
   \begin{tabular}{lccc}
      \tabularnewline \midrule \midrule
      Dependent Variables:              & InvEff (Abs)   & Positive Inv Resid & Negative Inv Resid\\  
      Model:                            & (1)            & (2)                & (3)\\  
      \midrule
      \emph{Variables}\\
      Treat x Post                      & -0.0007        & 0.0024             & -0.0017\\   
                                        & (-0.1365)      & (0.4322)           & (-0.5519)\\   
   \rowcolor{Gainsboro!60} % \rowcolor{lightgray}
      Treat x Post x High Customer Size & -0.0201$^{**}$ & 0.0150$^{*}$       & 0.0050\\   
   \rowcolor{Gainsboro!60} % \rowcolor{lightgray}
                                        & (-2.298)       & (1.708)            & (1.133)\\   
      Seg Sales Growth                  & 0.0063$^{**}$  & 0.0005             & -0.0068$^{*}$\\   
                                        & (1.971)        & (0.2658)           & (-1.884)\\   
      Customer Sales Growth             & 0.0031$^{**}$  & -0.0030$^{***}$    & $-8.35\times 10^{-5}$\\    
                                        & (2.465)        & (-2.686)           & (-0.2635)\\   
      Seg Size                          & 0.0402         & -0.0586$^{**}$     & 0.0185$^{**}$\\   
                                        & (1.350)        & (-2.016)           & (2.280)\\   
      Seg Relative Size                 & 0.1154         & -0.1867            & 0.0713$^{*}$\\   
                                        & (0.6676)       & (-1.039)           & (1.657)\\   
      Industry q                        & -0.0045        & 0.0052             & -0.0007\\   
                                        & (-0.2777)      & (0.3087)           & (-0.1512)\\   
      Customer Size                     & 0.0053         & -0.0044            & -0.0009\\   
                                        & (1.365)        & (-1.158)           & (-0.9041)\\   
      \midrule
      \emph{Fixed-effects}\\
      Firm-segment                      & Yes            & Yes                & Yes\\  
      Industry-Year                     & Yes            & Yes                & Yes\\  
      \midrule
      \emph{Fit statistics}\\
      Observations                      & 4,103          & 4,103              & 4,103\\  
      R$^2$                             & 0.80278        & 0.77339            & 0.88736\\  
      Adjusted R$^2$                    & 0.76759        & 0.73297            & 0.86726\\  
      \midrule \midrule
      \multicolumn{4}{l}{\emph{Clustered (Firm-segment) co-variance matrix, t-stats in parentheses}}\\
      \multicolumn{4}{l}{\emph{Signif. Codes: ***: 0.01, **: 0.05, *: 0.1}}\\
   \end{tabular}

   \caption*{\footnotesize
This table presents DiD regression results for the cross-sectional effect of customer disclosures on supplier-segment investments. The cross-sectioning variable \emph{High Customer Size} equals one if customer size is above the median and zero otherwise. The sample split is based on pre-period characteristics. All variable definitions are in \cref{sec:var_def}. The unit of observation is the supplier segment-customer segment-year. I include firm-segment and industry-year fixed effects following \textcite{chen-etal-2018-incorrect} and \textcite{chen-etal-2022-use}. $t$-statistics are reported in parentheses. Standard errors are clustered at the firm-segment level. All continuous variables are winsorized at the 1\% and 99\% levels.
   }
\end{table}

\clearpage
\begin{table}[htbp]
   \caption{\label{tbl:xsect_seg_02_hhi} (Segment-level Analysis) Customer Disclosures and Supplier-Segment Investment: The Role of Supplier-Segment Industry Concentration}
   \centering
   \footnotesize
   \renewcommand*{\arraystretch}{1.05}
   \begin{tabular}{lccc}
      \tabularnewline \midrule \midrule
      Dependent Variables:             & InvEff (Abs)   & Positive Inv Resid & Negative Inv Resid\\  
      Model:                           & (1)            & (2)                & (3)\\  
      \midrule
      \emph{Variables}\\
      Treat x Post                     & 0.0042         & 0.0014             & -0.0032\\   
                                       & (0.4999)       & (0.2665)           & (-1.162)\\   
   \rowcolor{Gainsboro!60} % \rowcolor{lightgray}
      Treat x Post x High Supp Seg HHI & -0.0222$^{**}$ & 0.0130$^{*}$       & 0.0024\\   
   \rowcolor{Gainsboro!60} % \rowcolor{lightgray}
                                       & (-2.165)       & (1.715)            & (0.4860)\\   
      Seg Sales Growth                 & 0.0220$^{*}$   & -0.0070            & -0.0092\\   
                                       & (1.775)        & (-0.9440)          & (-1.364)\\   
      Customer Sales Growth            & 0.0114         & -0.0013            & -0.0029$^{***}$\\   
                                       & (1.359)        & (-1.053)           & (-3.326)\\   
      Seg Size                         & 0.0290         & -0.0479$^{***}$    & 0.0266$^{**}$\\   
                                       & (1.030)        & (-2.820)           & (2.572)\\   
      Seg Relative Size                & -0.0196        & 0.0160             & 0.0186\\   
                                       & (-0.3129)      & (0.3604)           & (0.5297)\\   
      Industry q                       & 0.0208         & -0.0066            & -0.0048\\   
                                       & (1.137)        & (-0.5022)          & (-0.5555)\\   
      Customer Size                    & 0.0084         & -0.0009            & 0.0006\\   
                                       & (1.228)        & (-1.145)           & (1.419)\\   
      \midrule
      \emph{Fixed-effects}\\
      Firm-segment                     & Yes            & Yes                & Yes\\  
      Industry-Year                    & Yes            & Yes                & Yes\\  
      \midrule
      \emph{Fit statistics}\\
      Observations                     & 2,531          & 2,531              & 2,531\\  
      R$^2$                            & 0.81736        & 0.83517            & 0.90088\\  
      Adjusted R$^2$                   & 0.78862        & 0.80923            & 0.88528\\  
      \midrule \midrule
      \multicolumn{4}{l}{\emph{Clustered (Firm-segment) co-variance matrix, t-stats in parentheses}}\\
      \multicolumn{4}{l}{\emph{Signif. Codes: ***: 0.01, **: 0.05, *: 0.1}}\\
   \end{tabular}

   \caption*{\footnotesize
This table presents DiD regression results for the cross-sectional effect of customer disclosures on supplier-segment investments. The cross-sectioning variable \emph{High Supp Seg HHI} equals one for supplier segments operating in three-digit SIC industries with above-median concentration and zero otherwise. The sample split is based on pre-period characteristics. All variable definitions are in \cref{sec:var_def}. The unit of observation is the supplier segment-customer segment-year. I include firm-segment and industry-year fixed effects following \textcite{chen-etal-2018-incorrect} and \textcite{chen-etal-2022-use}. $t$-statistics are reported in parentheses. Standard errors are clustered at the firm-segment level. All continuous variables are winsorized at the 1\% and 99\% levels.
   }
\end{table}

\clearpage
% \input{tables/xsect_seg_03_s16_91_cinfo}

% consequences of over-investment
\begin{table}[htbp]
   \caption{\label{tbl:conseq_02_seg_roa} (Segment-level Analysis) Consequences of Supplier-Segment Investment: Future Segment Profitability}
   \centering
   \footnotesize
   \begin{adjustbox}{width = \textwidth, center}
      \renewcommand*{\arraystretch}{1.05}
      \begin{tabular}{lccc}
         \tabularnewline \midrule \midrule
         Dependent Variables:                        & Segment ROA (t+1) & Segment ROA (t+2) & Segment ROA (t+3)\\  
         Model:                                      & (1)               & (2)               & (3)\\  
         \midrule
         \emph{Variables}\\
   \rowcolor{Gainsboro!60} % \rowcolor{lightgray}
         Post $\times$ Positive Investment Residual  & -0.5634           & -0.8549$^{**}$    & -0.9430$^{*}$\\   
   \rowcolor{Gainsboro!60} % \rowcolor{lightgray}
                                                     & (-1.397)          & (-2.475)          & (-1.791)\\   
         Post                                        & 0.0774            & -0.0822           & -0.0048\\   
                                                     & (1.243)           & (-1.038)          & (-0.0358)\\   
         Positive Investment Residual                & 0.4564$^{*}$      & -0.0902           & 0.3241\\   
                                                     & (1.896)           & (-0.4782)         & (1.179)\\   
         log(Total Assets)                           & -0.2204$^{**}$    & -0.1108           & -0.1640\\   
                                                     & (-2.343)          & (-1.281)          & (-1.183)\\   
         Seg Sales Growth                            & -0.0163           & 0.0141            & 0.0300$^{***}$\\   
                                                     & (-0.7993)         & (1.068)           & (2.809)\\   
         Customer Sales Growth                       & -0.0324           & -0.0886           & -0.0899\\   
                                                     & (-0.6725)         & (-1.381)          & (-0.8730)\\   
         Seg Size                                    & -0.0386           & 0.0486            & 0.0978\\   
                                                     & (-0.6770)         & (0.7080)          & (0.8592)\\   
         Seg Relative Size                           & 0.0619            & -0.1714           & -0.4179\\   
                                                     & (0.3435)          & (-0.7017)         & (-1.095)\\   
         Industry q                                  & -0.1065$^{**}$    & 0.0643            & -0.0100\\   
                                                     & (-2.062)          & (1.533)           & (-0.2290)\\   
         \midrule
         \emph{Fixed-effects}\\
         Firm                                        & Yes               & Yes               & Yes\\  
         Year                                        & Yes               & Yes               & Yes\\  
         \midrule
         \emph{Fit statistics}\\
         Observations                                & 447               & 200               & 179\\  
         R$^2$                                       & 0.61670           & 0.62720           & 0.63877\\  
         Adjusted R$^2$                              & 0.38507           & 0.41585           & 0.44570\\  
         \midrule \midrule
         \multicolumn{4}{l}{\emph{Clustered (Firm) co-variance matrix, t-stats in parentheses}}\\
         \multicolumn{4}{l}{\emph{Signif. Codes: ***: 0.01, **: 0.05, *: 0.1}}\\
      \end{tabular}
   \end{adjustbox}

   \caption*{\footnotesize
This table presents OLS regression results for the association between positive segment investment residuals and segment ROA in subsequent years. A higher value of \emph{Positive Investment Residual} indicates investment above the level predicted by the segment sales-growth model. The outcome variable \emph{Segment ROA} is segment profit scaled by lagged segment assets in years $t+1$ through $t+3$. All variable definitions are in \cref{sec:var_def}. The unit of observation is the segment-year. I include firm and year fixed effects. $t$-statistics are reported in parentheses. Standard errors are clustered at the firm level. All continuous variables are winsorized at the 1\% and 99\% levels.
   }
\end{table}

\clearpage

% additional analysis
\begin{table}[htbp]
   \caption{\label{tbl:addl_adv_sga} (Firm-level Analysis) Customer Disclosures and Suppliers' Advertising and SG\&A Expenditures}
   \centering
   \scriptsize
   \begin{adjustbox}{width = 0.84\textwidth, center}
      \renewcommand*{\arraystretch}{0.85}
      \begin{tabular}{lcccc}
         \tabularnewline \midrule \midrule
         Dependent Variables: & \multicolumn{2}{c}{Advertising Expense} & \multicolumn{2}{c}{SG\&A Expense}\\
         Model:                     & (1)                    & (2)                   & (3)             & (4)\\  
         \midrule
         \emph{Variables}\\
   \rowcolor{Gainsboro!60} % \rowcolor{lightgray}
         \% Treat x Post            & 0.0135$^{**}$          & 0.0156$^{**}$         & 0.0224$^{**}$   & 0.0225$^{**}$\\   
   \rowcolor{Gainsboro!60} % \rowcolor{lightgray}
                                    & (2.197)                & (2.463)               & (2.492)         & (2.535)\\   
         log(Total Assets)          & -0.0126$^{**}$         & -0.0125$^{**}$        & -0.0390$^{***}$ & -0.0397$^{***}$\\   
                                    & (-2.270)               & (-2.334)              & (-2.629)        & (-2.697)\\   
         Market-to-Book             & $-4.79\times 10^{-5}$  & $1.65\times 10^{-5}$  & -0.0018         & -0.0017\\   
                                    & (-0.7646)              & (0.2949)              & (-1.034)        & (-0.9762)\\   
         Cashflow                   & 0.0180                 & 0.0194                & -0.1139$^{***}$ & -0.1149$^{***}$\\   
                                    & (0.9725)               & (1.115)               & (-2.995)        & (-3.053)\\   
         CapEx                      & -0.0094                & -0.0107               & -0.0101         & -0.0070\\   
                                    & (-0.6546)              & (-0.7294)             & (-0.2059)       & (-0.1392)\\   
         NonCapEx                   & 0.0084                 & 0.0134                & -0.0012         & -0.0034\\   
                                    & (1.295)                & (1.450)               & (-0.1396)       & (-0.3810)\\   
         Tangibility                & 0.0417$^{*}$           & 0.0169                & -0.0651         & -0.0612\\   
                                    & (1.715)                & (0.7914)              & (-0.5544)       & (-0.5224)\\   
         Cash                       & 0.0109                 & -0.0023               & 0.0691          & 0.0758\\   
                                    & (0.4118)               & (-0.0813)             & (0.8697)        & (0.9553)\\   
         Leverage                   & 0.0170                 & 0.0157                & 0.0309          & 0.0304\\   
                                    & (0.8615)               & (0.7575)              & (0.4434)        & (0.4304)\\   
         Dividend                   & 0.0075                 & 0.0062                & -0.0038         & -0.0036\\   
                                    & (1.191)                & (0.9450)              & (-0.3699)       & (-0.3451)\\   
         External Financing         & -0.0002                & -0.0003               & 0.0009          & 0.0011\\   
                                    & (-0.5710)              & (-0.8927)             & (0.7658)        & (0.8704)\\   
         N Segment                  & -0.0237$^{*}$          & -0.0278$^{***}$       & 0.0114$^{*}$    & 0.0127$^{**}$\\   
                                    & (-1.873)               & (-3.601)              & (1.846)         & (2.017)\\   
         Speed of Profit Adjustment &                        & -0.0141               &                 & 0.0323$^{*}$\\   
                                    &                        & (-1.464)              &                 & (1.860)\\   
         Concentration Ratio        &                        & -0.0458               &                 & 0.1565\\   
                                    &                        & (-0.6668)             &                 & (0.6184)\\   
         Seg Earnings Persistence   &                        & -0.0027               &                 & 0.0080\\   
                                    &                        & (-0.5730)             &                 & (0.5731)\\   
         Seg Industry Diversity     &                        & -0.0689$^{**}$        &                 & 0.0067\\   
                                    &                        & (-2.433)              &                 & (0.2054)\\   
         \midrule
         \emph{Fixed-effects}\\
         Firm                       & Yes                    & Yes                   & Yes             & Yes\\  
         Year                       & Yes                    & Yes                   & Yes             & Yes\\  
         \midrule
         \emph{Fit statistics}\\
         Observations               & 172                    & 172                   & 824             & 824\\  
         R$^2$                      & 0.96607                & 0.96943               & 0.94525         & 0.94563\\  
         Adjusted R$^2$             & 0.93762                & 0.94127               & 0.91165         & 0.91157\\  
         \midrule \midrule
         \multicolumn{5}{l}{\emph{Clustered (Firm) co-variance matrix, t-stats in parentheses}}\\
         \multicolumn{5}{l}{\emph{Signif. Codes: ***: 0.01, **: 0.05, *: 0.1}}\\
      \end{tabular}
   \end{adjustbox}

   \caption*{\footnotesize
This table presents DiD regression results for the effect of customer disclosures on supplier advertising and selling, general, and administrative (SG\&A) expenses. All variable definitions are in \cref{sec:var_def}. The unit of observation is the firm-year. I include firm and year fixed effects. $t$-statistics are reported in parentheses. Standard errors are clustered at the firm level. All continuous variables are winsorized at the 1\% and 99\% levels.
   }
\end{table}

\clearpage

% appendix example
% \subsection{Examples}
\section{Illustrative Examples}\label{illustrative-examples}

\subsection{Segment Disclosure Example}\label{sec:example_sfas_131}

Navistar International Corporation (CIK 0000808450) provides an
illustrative example of the impact of SFAS 131 implementation. In its
1998 10-K filing, under SFAS 14, Navistar disclosed only two segments.
However, following the adoption of SFAS 131, the company's 1999 10-K
filing revealed a more granular segment structure. Specifically,
Navistar disaggregated its previously reported Manufacturing Operations
segment into two distinct segments: Truck and Engine. Moreover, Navistar
provided restated segment-level data for previous years, demonstrating
how the breakdown would have appeared had SFAS 131 been in effect
earlier. This restatement allows potential suppliers to observe sales
growth of individual segments and to identify specific growth
opportunities, thereby illustrating how enhanced disclosures can
influence the competitive landscape of supplier industries.

\begin{figure}[H]
\centering
% Vector callout coordinates reproduce JIAR slide 9/41.
\begingroup
\setlength{\unitlength}{0.008\textwidth}
\begin{picture}(100,74.74)
\put(0,0){\includegraphics[width=0.8\textwidth]{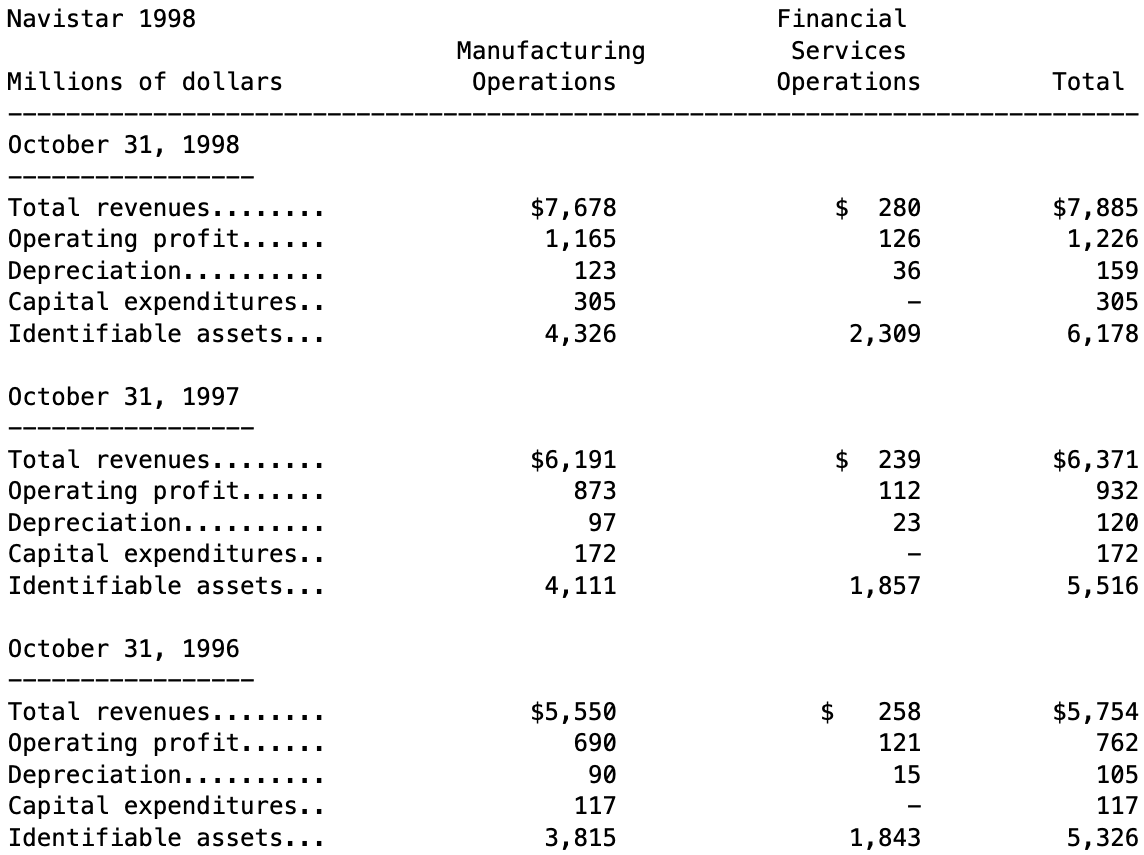}}
\color{red}
\linethickness{1.2pt}
\put(37.79,0){\framebox(21.05,74.74){}}
\end{picture}
\endgroup
\caption{Navistar's segment disclosure in 1998 (SFAS 14)}
\end{figure}

\begin{figure}[H]
\centering
% Vector callout coordinates reproduce JIAR slide 9/41.
\begingroup
\setlength{\unitlength}{0.008\textwidth}
\begin{picture}(100,80.19)
\put(0,0){\includegraphics[width=0.8\textwidth]{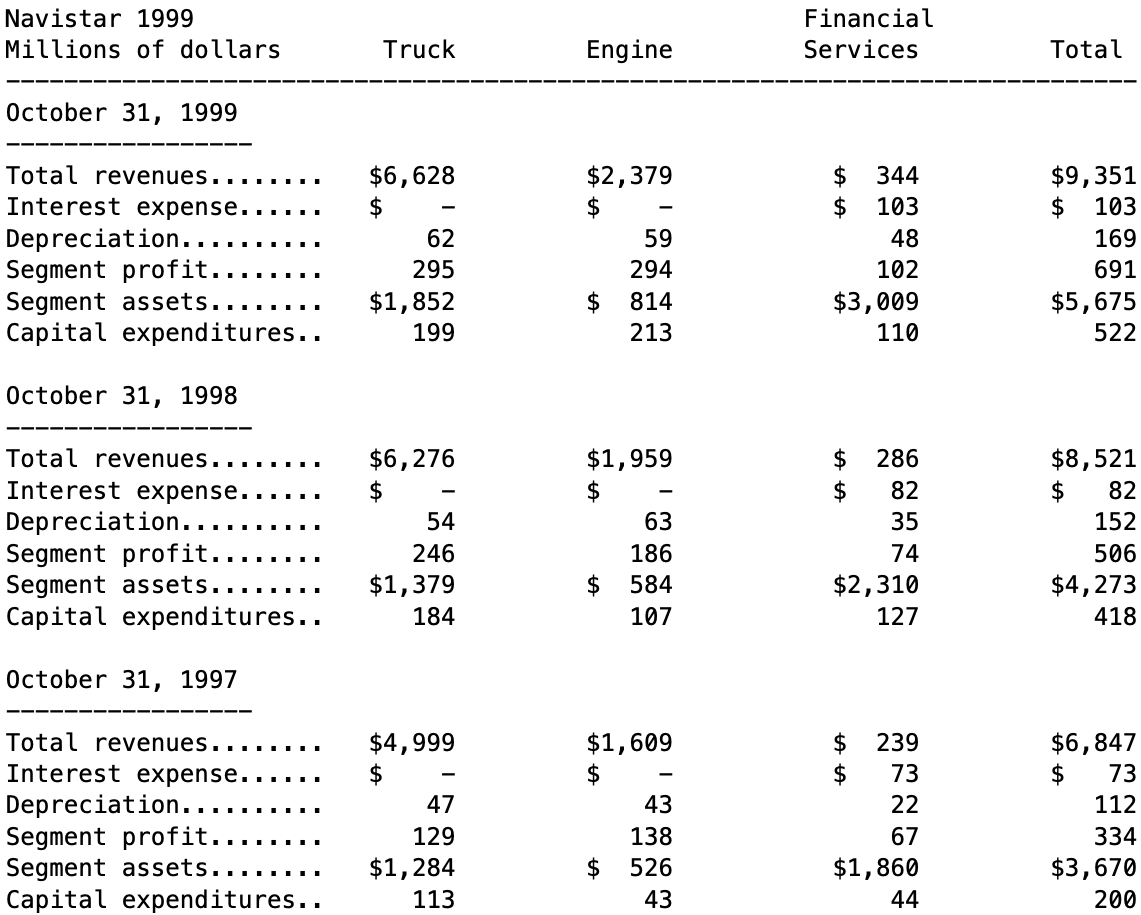}}
\color{red}
\linethickness{1.2pt}
\put(28.75,0){\framebox(35.89,80.19){}}
\end{picture}
\endgroup
\caption{Navistar's segment disclosure in 1999 (SFAS 131)}
\end{figure}

\subsection{Treatment and Post Variables Illustration: Increase in
Customer Disclosure}\label{sec:example_treat_post}

This example illustrates how the treatment and post variables are
implemented in this study. Consider a supplier company, HMT Technology
Corporation (CIK 0001005967). The supplier has three major customers:
Iomega Corporation, Maxtor Corporation, and Western Digital Corporation.
Among these customers, only Iomega Corporation increased its segment
disclosures following the adoption of SFAS 131.

The treatment variable for this supplier is calculated as 1/3,
representing the fraction of its major customers that expanded their
disclosures relative to the major customer base. The post variable
becomes 1 in fiscal year 1999 for this supplier, indicating the period
after SFAS 131 implementation. \cref{tbl:ex_treat_post} demonstrates how
these variables evolve over time.

\begin{figure}[H]
\centering
\includegraphics[width=0.8\textwidth,height=\textheight]{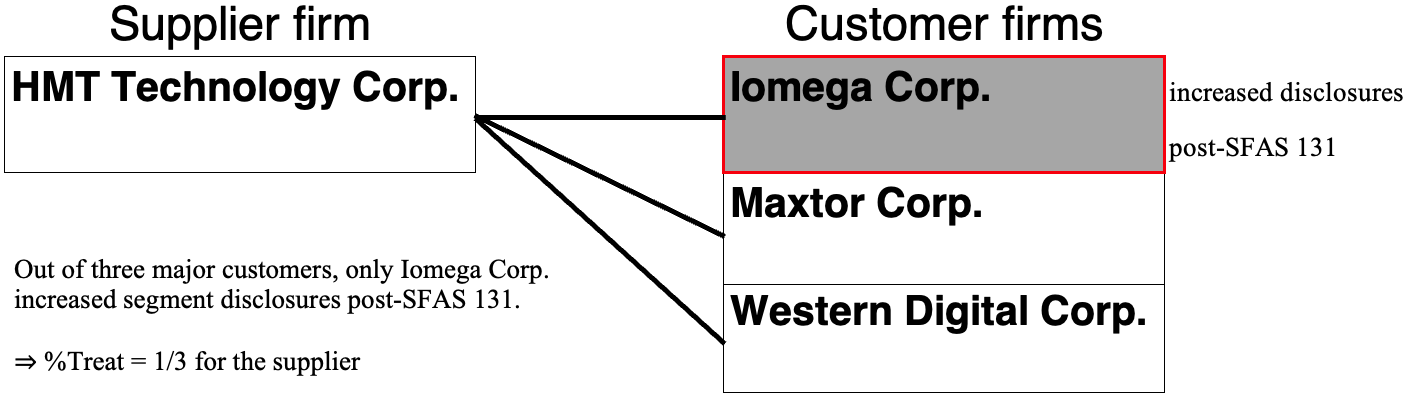}
\caption{\%Treat Calculation for HMT Technology Corp.}
\end{figure}

\begin{table}[!htbp] \centering 
  \caption{Treatment and Post Variables for HMT Technology Corporation}
  \label{tbl:ex_treat_post} % Replace 'mylabel' with a suitable label
  \begingroup
  \centering
\begin{tabular}{l|rrrr|}
\toprule

Year                   & 1997 & 1998 & 1999 & 2000 \\
\midrule
$\% Treat$             & 1/3  & 1/3  & 1/3  & 1/3  \\
$Post$                 & 0    & 0    & 1    & 1    \\
\midrule
$\% Treat \times Post$ & 0    & 0    & 1/3  & 1/3  \\
\midrule \midrule
\end{tabular}
\par\endgroup
%    \caption*{\footnotesize
% abc
%    }
\end{table}

The treatment variable remains constant at 1/3 throughout the period,
reflecting the stable proportion of customers that increased their
segment disclosures. The post variable switches from 0 to 1 in 1999,
marking the implementation of SFAS 131 for this supplier. Consequently,
the \(\%Treat \times Post\) interaction term takes the value 1/3
starting in 1999, capturing the effect of customer disclosure changes in
the post-SFAS 131 period.

\subsection{Segment-Segment Matching between Suppliers and
Customers}\label{sec:example_seg_seg_matching}

The Bureau of Economic Analysis (BEA) Input-Output (IO) data provide
valuable insights into the interdependencies between different
industries in the economy. These data reveal the average amount of input
required from various industries to produce one dollar of output in a
specific industry.\footnote{In this study, an industry producing an
  output in the BEA IO data is considered a customer industry, while an
  industry producing an input is considered a supplier industry.}

I use these data to match segments of customer companies to those of
supplier companies. Specifically, given a segment of a customer company,
I rank the segments of its supplier company based on how closely these
segments are economically related, as proxied by the Input-Output data.
This approach allows me to identify the most relevant supplier segment
for each customer segment, based on the strength of their economic
relationships.

To illustrate this segment-segment matching process, consider the
example of Safety Components International, Inc.~(CIK 0000918964) and
its customer, TRW, Inc.~(CIK 0000100030). Safety Components
International has two segments: Automotive Airbags and Defense. TRW
operates across two segments: Automotive and Space \& Defense.

The matching process begins by identifying an industry code (BEA IO
code) for each segment based on its SIC industry code. For instance:

\begin{enumerate}
\def\labelenumi{\arabic{enumi}.}
\tightlist
\item
  Safety Components International (supplier)

  \begin{itemize}
  \tightlist
  \item
    Automotive Airbags segment: BEA IO code 3361MV (Motor vehicles)
  \item
    Defense segment: BEA IO code 332 (Fabricated metal products)
  \end{itemize}
\item
  TRW (customer)

  \begin{itemize}
  \tightlist
  \item
    Automotive segment: BEA IO code 3361MV (Motor vehicles)
  \item
    Space \& Defense segment: BEA IO code 3364OT (Other transportation
    equipment)
  \end{itemize}
\end{enumerate}

Next, I use the BEA IO data to determine the input requirements for each
customer segment. According to the 1998 BEA IO data, to produce one
dollar of output in the Motor vehicles industry (IO code 3361MV),
\$0.231 of input from the Motor vehicles industry itself (IO code
3361MV) and \$0.058 of input from the Fabricated metal products industry
(IO code 332) are required. Therefore, I match the Automotive segment of
the customer company to the Automotive Airbags segment of the supplier
company over the Defense segment of the supplier.

\begin{figure}[H]
\centering
\includegraphics[width=0.8\textwidth,height=\textheight]{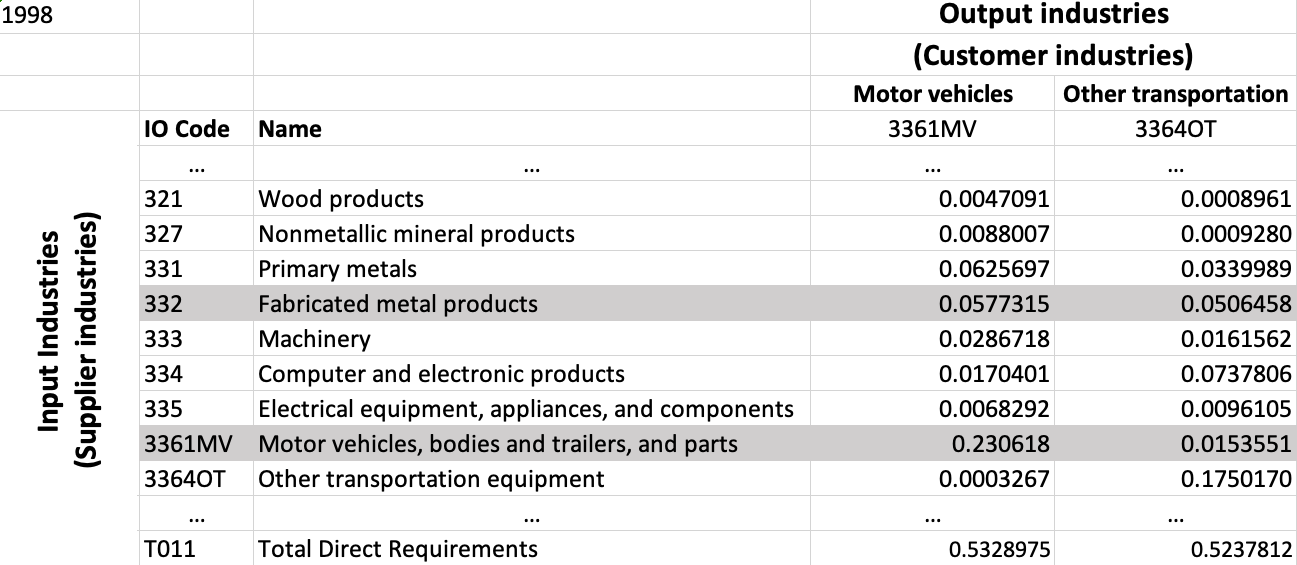}
\caption{Excerpt from the BEA Direct Requirements table}\label{fig:bea_table}
\end{figure}

\begin{figure}[H]
\centering
\begin{subfigure}{\textwidth} \centering
\includegraphics[width=0.8\textwidth,height=\textheight]{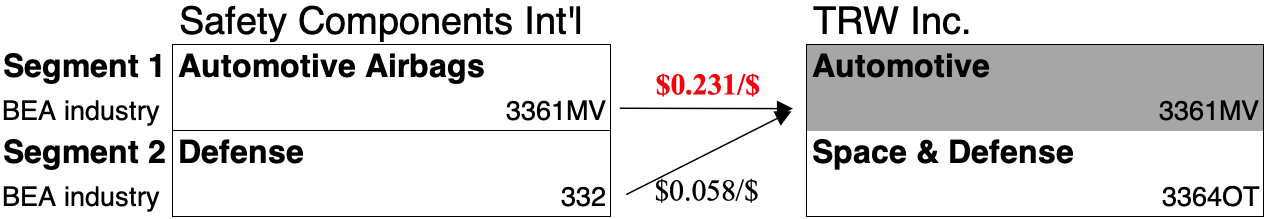}
\caption{Segment Matching: Customer's Automotive Segment to Supplier's Automotive Airbags Segment}\label{fig:seg_seg_match_01}
\end{subfigure}

\bigskip

\begin{subfigure}{\textwidth} \centering
\includegraphics[width=0.8\textwidth,height=\textheight]{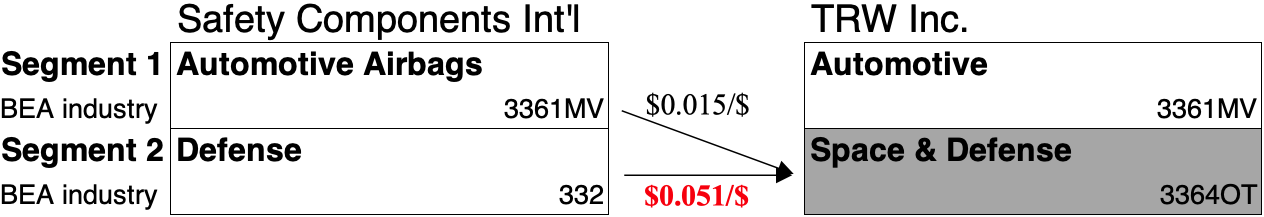}
\caption{Segment Matching: Customer's Space \& Defense Segment to Supplier's Defense Segment}\label{fig:seg_seg_match_02}
\end{subfigure}

\caption{Example of Matching Customer's Segments to Supplier's Segments}\label{fig:seg_seg_match_whole}
\end{figure}

Following the same methodology, the customer's Space \& Defense segment
is matched to the supplier's Defense segment. See
\Cref{fig:bea_table,fig:seg_seg_match_whole}.

\newpage

\subsection{Hand-Collecting Major Customer Information from 10-Ks: An
Example}\label{sec:example_hand_customer}

The WRDS Supply Chain database reports that Kitty Hawk, Inc.~(CIK
0000932110) had three major customers in 1996: General Motors, USPS, and
Burlington Northern Santa Fe (BNSF). Interestingly, the database listed
no customers for 1997 and only one (USPS) for 1998. However, an analysis
of the company's 10-K filings from 1996 to 1998 reveals a consistent
presence of three major customers throughout this period. While Kitty
Hawk chose not to explicitly name these customers in later filings,
their identities can be inferred by comparing the proportions of sales
across the years.

The 1996 10-K filing disclosed that GM, USPS, and Burlington represented
41.0\%, 14.9\%, and 10.9\% of total revenues, respectively. The 1997
10-K, while not naming specific customers, reported that one customer
accounted for approximately 41\% and 18\% of revenues in 1996 and 1997;
another for about 15\% and 18\%; and a third for approximately 12\% and
18\% in the same years. By analyzing these sales proportions, we can
infer that GM, USPS, and Burlington remained Kitty Hawk's major
customers throughout this period, despite the non-disclosures in these
filings.

\begin{figure}[H]
\centering
\includegraphics[width=0.9\textwidth,height=\textheight]{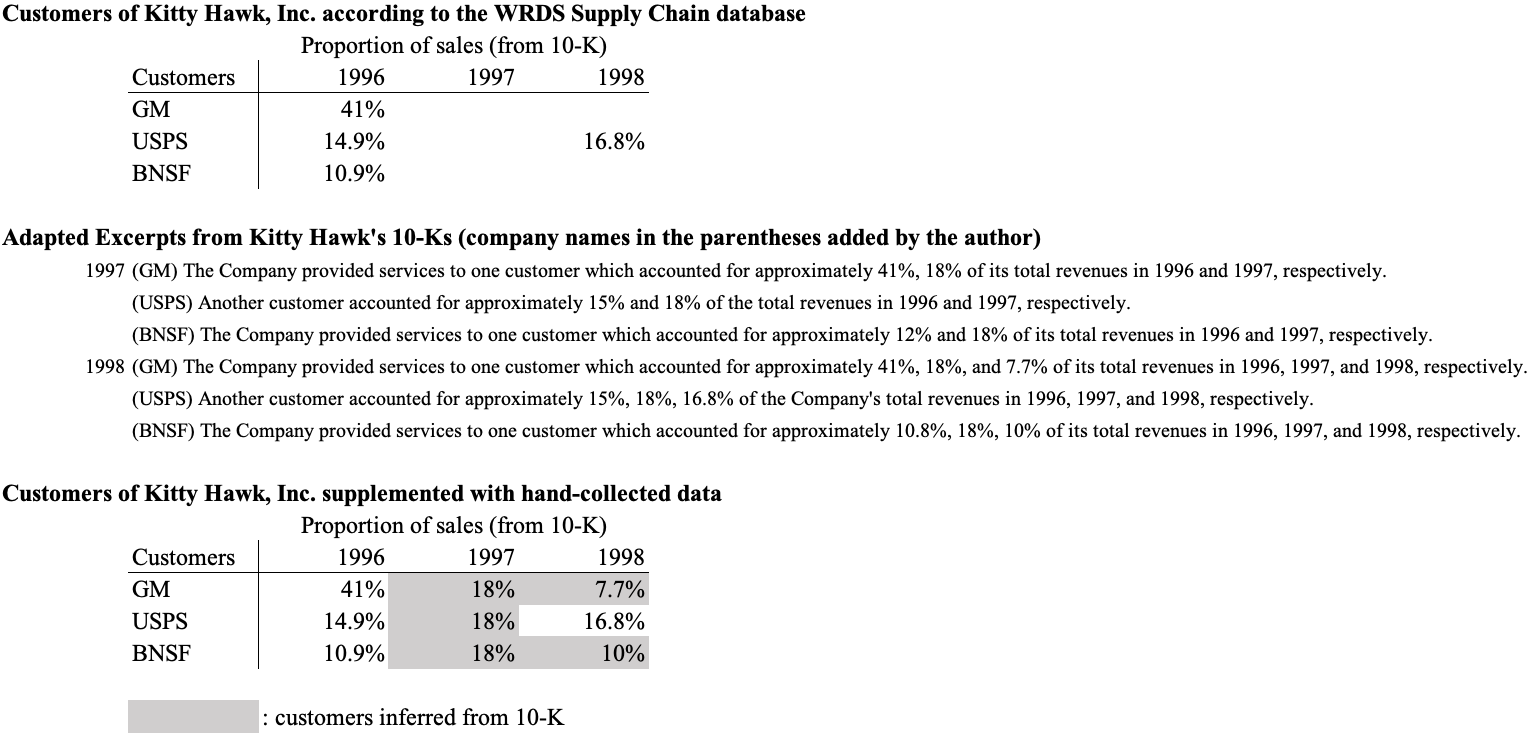}
\caption{Example of Major Customer Information Hand-Collected from Kitty Hawk's 10-Ks}\label{fig:example_kitty}
\end{figure}

\cref{fig:example_kitty} illustrates the hand-collected major customer
data from Kitty Hawk's 10-K reports, providing a clear visualization of
the company's customer relationships over time.

% Internet Appendix
\clearpage
\setcounter{section}{0}  % Reset section counter
\renewcommand{\thesection}{IA.\arabic{section}}  % Change section numbering format
\setcounter{table}{0}  % Reset table counter
\renewcommand{\thetable}{IA.\arabic{table}}  % Change table numbering format
\counterwithin*{table}{section}  % Remove section dependency if it exists

% Remove "Subsection" from subsections
\titleformat{\subsection}
{\normalfont\large\bfseries}{\thesubsection}{1em}{}

% Redefine section format for Internet Appendix
\titleformat{\section}
{\normalfont\Large\bfseries}{IA.\arabic{section}}{1em}{}

\centerline{\Large \bf Internet Appendix to} \vspace{2em} 
\centerline{\Large ``Do Customer Disclosures Affect Suppliers' Internal Capital Allocation Decisions?"}

%\section*{Internet Appendix to \\
%	``Do Customer Disclosures Affect Suppliers' Internal Capital Allocation Decisions?"}
\addcontentsline{toc}{section}{Internet Appendix}

\vspace{2em}  % Add some vertical space

\begin{center}
	\Large\textbf{Table of Contents}
\end{center}

\vspace{1em}  % Add some vertical space

\begin{enumerate}[leftmargin=*, label=IA.\arabic*., font=\normalfont\large, itemsep=0.5em]
	\item The Effect of Customer Disclosures on Their Competitive Environment
	\item Customer Competition Changes and Supplier Internal Capital Reallocation
	% Add more items as needed
\end{enumerate}

\clearpage

\begin{table}[htbp]
   \caption{\label{tbl:ia_customer_competition} (Firm-level Analysis) Customer Disclosures and Competition in Customer Product Markets}
   \centering
   \scriptsize
   \begin{adjustbox}{width = \textwidth, center}
      \renewcommand*{\arraystretch}{1}
      \begin{tabular}{lcccccccc}
         \tabularnewline \midrule \midrule
         Dependent Variables: & \multicolumn{4}{c}{N Rivals (Hoberg-Phillips)} & \multicolumn{4}{c}{Textual Similarities (Hoberg-Phillips)}\\
         Model:                     & (1)      & (2)           & (3)      & (4)           & (5)      & (6)            & (7)     & (8)\\  
         \midrule
         \emph{Variables}\\
   \rowcolor{Gainsboro!60} % \rowcolor{lightgray}
         Treat x Post               & 5.220    & 2.191         & 4.489    & 0.8084        & 0.1839   & 0.0726         & 0.2237  & 0.0561\\   
   \rowcolor{Gainsboro!60} % \rowcolor{lightgray}
                                    & (0.9728) & (0.3904)      & (0.7932) & (0.1262)      & (0.9281) & (0.3344)       & (1.160) & (0.2531)\\   
         log(Total Assets)          &          & -18.08$^{**}$ &          & -18.47$^{**}$ &          & -0.8349$^{**}$ &         & -0.8785$^{**}$\\   
                                    &          & (-2.002)      &          & (-1.976)      &          & (-2.044)       &         & (-2.555)\\   
         Market-to-Book             &          & -0.4867       &          & -0.0623       &          & -0.0227        &         & -0.0026\\   
                                    &          & (-0.9259)     &          & (-0.1038)     &          & (-1.005)       &         & (-0.0946)\\   
         Cashflow                   &          & 29.62         &          & 31.24         &          & 0.9936         &         & 1.111\\   
                                    &          & (1.241)       &          & (1.322)       &          & (1.161)        &         & (1.415)\\   
         CapEx                      &          & 13.88         &          & -15.35        &          & 1.058          &         & -0.2040\\   
                                    &          & (0.8138)      &          & (-0.7444)     &          & (1.395)        &         & (-0.2471)\\   
         NonCapEx                   &          & -3.558        &          & -12.37        &          & 0.0700         &         & -0.3897\\   
                                    &          & (-0.5482)     &          & (-1.568)      &          & (0.2164)       &         & (-1.125)\\   
         Tangibility                &          & -1.553        &          & -34.11        &          & -0.0842        &         & -1.298\\   
                                    &          & (-0.0505)     &          & (-0.8275)     &          & (-0.0804)      &         & (-0.8229)\\   
         Cash                       &          & -42.11        &          & -53.28        &          & -1.593         &         & -2.424\\   
                                    &          & (-0.8406)     &          & (-0.9653)     &          & (-0.8319)      &         & (-1.199)\\   
         Leverage                   &          & 41.10         &          & 22.23         &          & 1.359          &         & 0.6224\\   
                                    &          & (1.242)       &          & (0.6387)      &          & (1.009)        &         & (0.4890)\\   
         Dividend                   &          & -8.167        &          & -13.22        &          & -0.2801        &         & -0.4006\\   
                                    &          & (-0.4976)     &          & (-0.7093)     &          & (-0.6010)      &         & (-0.7188)\\   
         External Financing         &          & 0.8081        &          & 0.9875        &          & 0.0362         &         & 0.0484$^{*}$\\   
                                    &          & (1.288)       &          & (1.335)       &          & (1.615)        &         & (1.893)\\   
         N Segment                  &          & 2.845         &          & 2.737         &          & 0.0427         &         & 0.0314\\   
                                    &          & (1.328)       &          & (1.419)       &          & (0.6125)       &         & (0.4836)\\   
         Speed of Profit Adjustment &          & 3.126         &          & -44.46$^{**}$ &          & 0.2105         &         & -1.634$^{**}$\\   
                                    &          & (0.4891)      &          & (-2.248)      &          & (0.8607)       &         & (-2.237)\\   
         Concentration Ratio        &          & 25.95         &          & -60.26        &          & 1.701          &         & -0.0474\\   
                                    &          & (0.4578)      &          & (-0.4178)     &          & (0.8270)       &         & (-0.0094)\\   
         Seg Earnings Persistence   &          & -0.2665       &          & -14.15        &          & 0.0018         &         & -0.4463\\   
                                    &          & (-0.5443)     &          & (-1.190)      &          & (0.0982)       &         & (-1.155)\\   
         Seg Industry Diversity     &          & 16.72         &          & 3.945         &          & 0.7146$^{*}$   &         & 0.1508\\   
                                    &          & (1.104)       &          & (0.2364)      &          & (1.864)        &         & (0.3372)\\   
         \midrule
         \emph{Fixed-effects}\\
         (Customer) Firm            & Yes      & Yes           & Yes      & Yes           & Yes      & Yes            & Yes     & Yes\\  
         Year                       & Yes      & Yes           &          &               & Yes      & Yes            &         & \\  
         Industry-Year              &          &               & Yes      & Yes           &          &                & Yes     & Yes\\  
         \midrule
         \emph{Fit statistics}\\
         Observations               & 711      & 674           & 711      & 674           & 740      & 702            & 740     & 702\\  
         R$^2$                      & 0.90282  & 0.90792       & 0.92614  & 0.93105       & 0.89387  & 0.90110        & 0.92321 & 0.92902\\  
         Adjusted R$^2$             & 0.87007  & 0.87196       & 0.85187  & 0.85222       & 0.85920  & 0.86380        & 0.84868 & 0.85103\\  
         \midrule \midrule
         \multicolumn{9}{l}{\emph{Clustered ((Customer) Firm) co-variance matrix, t-stats in parentheses}}\\
         \multicolumn{9}{l}{\emph{Signif. Codes: ***: 0.01, **: 0.05, *: 0.1}}\\
      \end{tabular}
   \end{adjustbox}

   \caption*{\footnotesize
This table presents DiD regression results for the effect of customer disclosures on customers' competitive environment. I measure competition using two measures based on the Text-based Network Industry Classifications (TNIC-3; see \textcite{hoberg-phillips-2010-product} and \textcite{hoberg-phillips-2016-textbased}). The \emph{N Rivals} measure is the number of rivals in the same TNIC-3 industry, and the \emph{Textual Similarities} measure is the textual similarity of product descriptions with those rivals. The sample consists of customer firms, and \emph{Treat} equals one when the customer itself expanded segment disclosure following SFAS 131. All other variable definitions are in \cref{sec:var_def}. The unit of observation is the customer firm-year. Columns (1), (2), (5), and (6) include customer-firm and year fixed effects. Columns (3), (4), (7), and (8) include customer-firm and industry-year fixed effects, where industries are defined at the two-digit SIC level. $t$-statistics are reported in parentheses. Standard errors are clustered at the customer-firm level. All continuous variables are winsorized at the 1\% and 99\% levels.
   }
\end{table}

\clearpage

\begin{table}[htbp]
   \caption{\label{tbl:ia_customer_competition_reallocation} (Firm-level Analysis) Competition in Customer Product Markets and Suppliers' Internal Capital Allocation}
   \centering
   \scriptsize
   \begin{adjustbox}{width = 0.88\textwidth, center}
      \renewcommand*{\arraystretch}{0.87}
      \begin{tabular}{lccccc}
         \tabularnewline \midrule \midrule
         Dependent Variable: & \multicolumn{5}{c}{Capital Allocation Efficiency}\\
         Model:                                          & (1)            & (2)            & (3)           & (4)           & (5)\\  
         \midrule
         \emph{Variables}\\
   \rowcolor{Gainsboro!60} % \rowcolor{lightgray}
         \% Treat x Post                                 & -0.0503$^{**}$ & -0.0512$^{**}$ & -0.0548$^{*}$ & -0.0540$^{*}$ & -0.0611\\   
   \rowcolor{Gainsboro!60} % \rowcolor{lightgray}
                                                         & (-2.129)       & (-2.017)       & (-1.832)      & (-1.662)      & (-1.534)\\   
   \rowcolor{Gainsboro!60} % \rowcolor{lightgray}
         \% Treat x Post x Customer Rival Change (Std.)  & 0.0041         & 0.0036         &               &               &   \\   
   \rowcolor{Gainsboro!60} % \rowcolor{lightgray}
                                                         & (0.1822)       & (0.1480)       &               &               &   \\   
   \rowcolor{Gainsboro!60} % \rowcolor{lightgray}
         \% Treat x Post x Mid Customer Rival Change     &                &                & -0.0119       & -0.0113       & 0.0064\\   
   \rowcolor{Gainsboro!60} % \rowcolor{lightgray}
                                                         &                &                & (-0.3479)     & (-0.3087)     & (0.1574)\\   
   \rowcolor{Gainsboro!60} % \rowcolor{lightgray}
         \% Treat x Post x High Customer Rival Change    &                &                & 0.0377        & 0.0305        & 0.0833$^{*}$\\   
   \rowcolor{Gainsboro!60} % \rowcolor{lightgray}
                                                         &                &                & (1.080)       & (0.8551)      & (1.764)\\   
         log(Total Assets)                               &                & 0.0427         &               & 0.0411        & 0.0929$^{*}$\\   
                                                         &                & (1.167)        &               & (1.110)       & (1.788)\\   
         Market-to-Book                                  &                & -0.0005        &               & -0.0005       & 0.0024\\   
                                                         &                & (-0.0834)      &               & (-0.0802)     & (0.3089)\\   
         Cashflow                                        &                & -0.0301        &               & -0.0312       & -0.0396\\   
                                                         &                & (-0.3543)      &               & (-0.3708)     & (-0.3449)\\   
         CapEx                                           &                & 0.0236         &               & 0.0235        & 0.0208\\   
                                                         &                & (0.2375)       &               & (0.2357)      & (0.1683)\\   
         NonCapEx                                        &                & -0.0013        &               & -0.0022       & -0.0713\\   
                                                         &                & (-0.0286)      &               & (-0.0485)     & (-1.288)\\   
         Tangibility                                     &                & -0.0376        &               & -0.0330       & -0.0660\\   
                                                         &                & (-0.2251)      &               & (-0.1969)     & (-0.3163)\\   
         Cash                                            &                & 0.0745         &               & 0.0704        & -0.0071\\   
                                                         &                & (0.5371)       &               & (0.5099)      & (-0.0456)\\   
         Leverage                                        &                & -0.0806        &               & -0.0770       & -0.1294\\   
                                                         &                & (-1.056)       &               & (-0.9921)     & (-1.149)\\   
         Dividend                                        &                & 0.0037         &               & 0.0062        & -0.0141\\   
                                                         &                & (0.0981)       &               & (0.1648)      & (-0.3929)\\   
         External Financing                              &                & 0.0007         &               & 0.0009        & -0.0005\\   
                                                         &                & (0.2465)       &               & (0.2931)      & (-0.1272)\\   
         N Segment                                       &                & 0.0121         &               & 0.0106        & 0.0019\\   
                                                         &                & (0.3094)       &               & (0.2700)      & (0.0408)\\   
         Speed of Profit Adjustment                      &                & 0.0195         &               & 0.0229        & 0.0827\\   
                                                         &                & (0.2493)       &               & (0.2892)      & (0.5555)\\   
         Concentration Ratio                             &                & -2.014$^{**}$  &               & -2.013$^{**}$ & 1.333\\   
                                                         &                & (-2.181)       &               & (-2.175)      & (0.8263)\\   
         Seg Earnings Persistence                        &                & 0.0325         &               & 0.0302        & -0.0112\\   
                                                         &                & (0.2662)       &               & (0.2460)      & (-0.0729)\\   
         Seg Industry Diversity                          &                & -0.0296        &               & -0.0330       & -0.0600\\   
                                                         &                & (-0.2384)      &               & (-0.2676)     & (-0.4471)\\   
         \midrule
         \emph{Fixed-effects}\\
         Firm                                            & Yes            & Yes            & Yes           & Yes           & Yes\\  
         Year                                            & Yes            & Yes            & Yes           & Yes           & \\  
         Industry-Year                                   &                &                &               &               & Yes\\  
         \midrule
         \emph{Fit statistics}\\
         Observations                                    & 857            & 857            & 857           & 857           & 857\\  
         R$^2$                                           & 0.55286        & 0.56283        & 0.55373       & 0.56343       & 0.73340\\  
         Adjusted R$^2$                                  & 0.30409        & 0.30053        & 0.30418       & 0.30017       & 0.38818\\  
         \midrule \midrule
         \multicolumn{6}{l}{\emph{Clustered (Firm) co-variance matrix, t-stats in parentheses}}\\
         \multicolumn{6}{l}{\emph{Signif. Codes: ***: 0.01, **: 0.05, *: 0.1}}\\
      \end{tabular}
   \end{adjustbox}

   \caption*{\footnotesize
This table presents DiD regression results testing whether suppliers alter internal capital allocation more strongly when their linked customers experience larger changes in product-market competition. The dependent variable is \emph{Capital Allocation Efficiency}. \emph{\% Treat} is the proportion of customers that changed their segment disclosure following SFAS 131 relative to the supplier's entire major-customer base. \emph{Customer Rival Change} is the supplier-level average of linked customers' changes in TNIC-3 rival counts from the pre-SFAS 131 period to the post-SFAS 131 period. The continuous measure is standardized; the middle- and high-change indicators identify the second and third terciles, with the lowest tercile omitted. All other variable definitions are in \cref{sec:var_def}. The unit of observation is the supplier firm-year. Columns (1) and (3) omit controls, whereas Columns (2), (4), and (5) include controls. Columns (1) through (4) include firm and year fixed effects. Column (5) includes firm and industry-year fixed effects, where industries are defined at the two-digit SIC level. $t$-statistics are reported in parentheses. Standard errors are clustered at the supplier-firm level. All continuous variables are winsorized at the 1\% and 99\% levels.
   }
\end{table}

\end{appendices}
\end{document}